\documentclass[journal]{IEEEtran}
\usepackage[caption=false]{subfig}
\usepackage{url}
\usepackage{graphicx}
\usepackage{hyperref}
\usepackage{balance}
\usepackage{makecell}
\usepackage{stfloats}
\usepackage{color,soul}
\usepackage{amsmath}
\usepackage{multirow}
\usepackage{hhline}
\usepackage{cite}
\usepackage{amssymb}
\usepackage{amsthm}
\usepackage{mathtools}
\usepackage{multicol}
\usepackage{comment}
\usepackage{adjustbox}
\usepackage{multirow}

\usepackage{booktabs, makecell, tabularx}
\newcolumntype{C}{>{\centering\arraybackslash}X} 

\usepackage{stfloats}

\usepackage{xcolor,colortbl}
\usepackage{enumitem}
\usepackage{blindtext}
\definecolor{Gray}{gray}{0.9}
\usepackage{tikz}
\usepackage{pifont}
\usepackage{wasysym}
\newcommand{\cmark}{\CIRCLE}
\newcommand{\xmark}{\Circle}%
\newcommand{\smark}{\LEFTcircle}

\newcolumntype{P}[1]{>{\centering\arraybackslash}p{#1}}
\newcolumntype{M}[1]{>{\centering\arraybackslash}m{#1}}

\begin{document}
\bstctlcite{IEEEexample:BSTcontrol}

\title{Applications of Generative AI (GAI) for \\Mobile and Wireless Networking: A Survey}

\author{Thai-Hoc Vu, Senthil~Kumar~Jagatheesaperumal, Minh-Duong~Nguyen, \\Nguyen Van Huynh, Sunghwan Kim, and Quoc-Viet Pham

\thanks{T.-H. Vu is with the Department of Electrical, Electronic and Computer Engineering, University of Ulsan, Republic of Korea (email: vuthaihoc1995@gmail.com). S. K. Jagatheesaperumal (co-first author) is with the Department of Electronics and Communication Engineering, Mepco Schlenk Engineering College, Sivakasi 626005, India (e-mail: senthilkumarj@mepcoeng.ac.in). M.-D. Nguyen is with the Department of Information Convergence Engineering, Pusan National University, Busan 46241, Republic of Korea (e-mail: duongnm@pusan.ac.kr). N. V. Huynh is with the Department of Electrical Engineering and Electronics, University of Liverpool, Liverpool, L69 3GJ, United Kingdom (e-mail: huynh.nguyen@liverpool.ac.uk). S. Kim (corresponding author) is with the School of Electronic Engineering, Kyonggi University, Republic of Korea (email: skim@kyonggi.ac.kr). Q.-V. Pham (senior author) is with the School of Computer Science and Statistics, Trinity College Dublin, Dublin 2, D02 PN40, Ireland (e-mail: viet.pham@tcd.ie).}

\thanks{This work was supported by the Research Program through the National Research Foundation of Korea under Grant NRF-2023R1A2C1003546.}

}

\maketitle

\begin{abstract}
The success of Artificial Intelligence (AI) in multiple disciplines and vertical domains in recent years has promoted the evolution of mobile networking and the future Internet toward an AI-integrated Internet-of-Things (IoT) era. Nevertheless, most AI techniques rely on data generated by physical devices (e.g., mobile devices and network nodes) or specific applications (e.g., fitness trackers and mobile gaming). Therefore, Generative AI (GAI), a.k.a. AI-generated content (AIGC), has emerged as a powerful AI paradigm; thanks to its ability to efficiently learn complex data distributions and generate synthetic data to represent the original data in various forms. This impressive feature is projected to transform the management of mobile networking and diversify the current services and applications provided. On this basis, this work presents a concise tutorial on the role of GAIs in mobile and wireless networking. In particular, this survey first provides the fundamentals of GAI and representative GAI models, serving as an essential preliminary to the understanding of GAI's applications in mobile and wireless networking. Then, this work provides a comprehensive review of state-of-the-art studies and GAI applications in network management, wireless security, semantic communication, and lessons learned from the open literature. Finally, this work summarizes the current research on GAI for mobile and wireless networking by outlining important challenges that need to be resolved to facilitate the development and applicability of GAI in this edge-cutting area.


\end{abstract}


\markboth{Journal of \LaTeX\ Class Files,~Vol.~X, No.~Y, Z~2024 - This work has been accepted for publication in IoT Journal}%
{Shell \MakeLowercase{\textit{et al.}}: Bare Demo of IEEEtran.cls for IEEE Journals}

\IEEEpeerreviewmaketitle
\section{Introduction} 

\label{sec:Introduction}
Generative Artificial Intelligence (GAI), a prominent branch of Artificial Intelligence (AI) responsible for AI-Generated Content (AIGC), has recently attracted significant attention from the scientific community to big tech giants and governments. The three most popular representations of GAI are ChatGPT, Bing, and Google Bard, which are Large Language Model (LLM)-based chatbots that can conversationally interact with humans and generate new data based on the input and contextual data~\cite{gai_google}. Different from conventional AI and Machine Learning (ML) systems that are only trained on a certain dataset to create mapping abstractions to make highly accurate decisions, GAI platforms do not just learn from the input dataset but also rely on the interaction of data to combine, modify, and create new information to reinforce its learning process. Driven by consumer interests, GAIs have recently emerged as a powerful tool for various uses. For example, in robotics control, Reinforcement Learning (RL) is a key technique for enabling robots to learn from their own experiences, which are then translated into appropriate movements and actions in real time. Through GAI interaction with RL, robots become more humanoid in generating new motions and movement skills without human intervention or initial knowledge provision~\cite{gai_example}. 
Another example is in the field of cybersecurity for the Internet-of-Things (IoT), Deep Generative Models (DGM), fundamental components behind GAI, help generate patterns and address critical challenges involving the lack of malicious samples and the rapid evolution of cyberattacks \cite{vu2023deep}. Consequently, adopting GAIs is projected to produce new ways to significantly improve current AI-enabled algorithms in mobile networking and the IoT.

\begin{figure}[t]
    \centering
    \includegraphics[width=\linewidth]{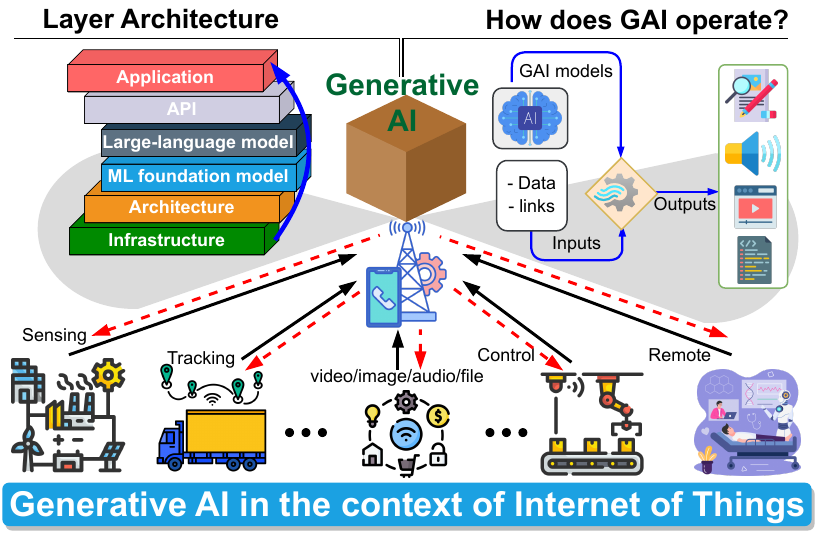}
    \caption{\color{black}Illustration of GAI in the context of wireless IoT networks.}
    \label{Fig:General_GAI}
\end{figure}

{\color{black}Fig.~\ref{Fig:General_GAI} illustrates GAI in the context of IoT, where its architecture consists of six layers, each dedicated to specific functions \cite{Epical2024Sep}. The foundation layer, \textit{Infrastructure}, encompasses computing resources based on GPUs, CPUs, or distributed systems, data storage solutions such as data lakes, databases, data warehouses, and network connections such as high-speed networks and cloud services.
\textit{Architecture}, a critical aspect of GAI's structural design, includes the optimal selection and configuration of neural networks for specific generative tasks. It also assesses the models' capability and efficiency in learning from data and generating new content, utilizing key components like Transformers and Generative Adversarial Networks (GANs). Herein, Transformers use Attention mechanisms to manage dependencies between input and output data, while GANs generate high-quality synthetic data. The third layer, \textit{Machine Learning (ML) foundation model}, involves building ML models trained on vast datasets to capture fundamental patterns, including training data, algorithms, and model evaluation, making GAI powerful in learning and generalizing from data. The LLM layer focuses on implementing LLMs using model architectures (e.g., Transformer, GPT, BERT, T5), training data with considerations of text corpora, diversity, and representation, and fine-tuning for task-specific adaptation or transfer learning. The \textit{Application Programming Interface (API)} layer provides standardized interfaces for accessing and utilizing GAI models, including Representational State Transfer Application APIs (RESTful APIs) for Hypertext Transfer Protocol (HTTP) activities, Graph Query Language (GraphQL) APIs for client requests and interactions, and software development kits (SDKs) for programming. Finally, the \textit{Application} layer bridges GAIs to real-world scenarios across various domains, showcasing the tangible benefits of GAI in improving efficiency and driving innovation, such as in chatbots, code generation, search engines, and cybersecurity. More specifically, when a typical user inputs queries like `\textit{What is generative AI?}'' or ``\textit{What are essential blocks of GAI?}'' to seek draft answers from GAI tools, the `grey' block, representing trained GAI models, combines with the `orange' block, which includes the system settings used as input to the GAI models, to generate output data in the form of text, audio, video, or code. The `green' block then processes the data generated by the GAI models to respond to the user's input. Thus, introducing GAI into not only mobile and wireless networking but also large-scale IoT will open exciting new research directions, paying the base to shift current network deployments close to the era of generative IoT \cite{Wen2024May}.} 

\subsection{State-of-the-Arts}
\label{sec:Introduction_Contributions}
AI techniques have become widespread in various research areas in mobile networking and the IoT, such as network management, wireless security, resource allocation, and edge services. In recent years, data-driven and AI-enabled approaches have established superior performance over conventional optimization-based designs, which are limited primarily by network scalabilities in handling massive connections and heterogeneous variations~\cite{pham2021intelligent}. Additionally, AI techniques enable efficient approximation of computationally challenging algorithms, to effectively overcome nonlinear and unknown patterns in wireless networks, and rapidly accelerate existing model-based algorithms.
On this basis, there have been several surveys of different AI paradigms and their applications for mobile networking, such as Deep Learning (DL)~\cite{mao2018deep, zhang2019deep}, Deep Reinforcement Learning (DRL)~\cite{luong2019applications, li2022applications}, Federated Learning (FL)~\cite{lim2020federated, nguyen2021federated}, and swarm intelligence~\cite{pham2021swarm}.
However, current AI-based approaches for mobile and wireless networking face several issues. Specifically, conventional AI algorithms usually require a large volume of labeled training data to facilitate better performance. Unfortunately, collecting and labeling training data in complex environments like wireless networks are costly, time-consuming, and require intensive human interventions. In addition, with an enormous number of diverse users, future wireless systems are highly dynamic and uncertain. It can significantly degrade the performance of conventional AI models, which are usually trained in a specific environment/condition. Although the distributed learning paradigm can help to facilitate the learning process of AI-based approaches, it still suffers heavily from scarcity of data required for efficient training \cite{le2023applications}. Moreover, due to the differences in local system characteristics, the distributed AI system suffers from the long-tailed distributed data phenomenon. This data imbalance results in a negative transfer toward the data cluster with a dominant size. To address all these practical challenges, GAI is considered an efficient solution thanks to its properties of domain adaptation, data generation, and abnormal detection. Consequently, several surveys in the literature focus on the applications of GAI in various domains such as Network Management (NetMan), Semantic Communication (SemCom), and security and privacy, as summarized in Table~\ref{tab:surveys}.

\begin{table*}[!ht]
\centering
\caption{Existing survey papers about GAI for mobile networking. \\~\cmark, \smark\, and \xmark, indicate that the topic is well-covered, partially covered, and uncovered, respectively.} 
\label{tab:surveys}
\begin{tabular}{|p{0.5cm}|p{0.6cm}|p{9cm}|p{0.5cm}|p{0.5cm}|p{1.0cm}|p{1.0cm}|p{1.0cm}|}
\hline
\textbf{Ref.} 
& \textbf{Year} 
& \textbf{Focus} 
& \textbf{ML}
& \textbf{GAI}
& \textbf{NetMan}
& \textbf{Security}
& \textbf{SemCom}
\\ \hline
\hline
\multirow{1}*{\cite{yang2019generative} }& 2019 & Autonomous wireless channel modeling without complex theoretical analysis. & \cmark  & \cmark  & \smark & \xmark & \xmark \\ \hline
\multirow{1}*{\cite{ayanoglu2022machine}} & 2022 & Analysis techniques and applications of GAN in anomaly detection. &   \cmark  & \cmark  & \xmark & \cmark & \xmark \\ \hline
\multirow{1}*{\cite{yang2022generative}} & 2022 & GAN for security and reliable trust management in real-time communications. & \smark  & \cmark & \xmark & \cmark & \xmark \\ \hline
\multirow{1}*{\cite{cao2023comprehensive}} & 2023 & AIGC for content creation using unimodal and multimodal large-scale models. & \cmark & \cmark & \smark & \xmark & \xmark \\ \hline
\multirow{1}*{\cite{zhang2023complete}}& 2023 & AIGC models for output types in specific tasks (text, images, and videos). & \cmark & \cmark & \smark & \xmark & \xmark \\ \hline
\multirow{1}*{\cite{gozalo2023chatgpt}} & 2023 & Impact of GAI on different industries and taxonomy of recent key models. & \smark & \cmark  & \smark & \xmark  & \xmark  \\ \hline
\multirow{1}*{\cite{sabuhi2021applications}} & 2021 & Training solutions for GAN with mode collapse, and gradient vanishing. &   \cmark  & \cmark  & \xmark & \xmark & \xmark \\ \hline
\multirow{1}*{\cite{jabbar2021survey}} & 2021 & Analyses of GANs' stability, and training solutions by assessing obstacles. &   \cmark  & \cmark  & \smark & \cmark & \xmark \\ \hline
\multirow{1}*{\cite{navidan2021generative}} & 2021 & Computer and communication interplay (control, security, and mobile services).  &   \cmark  & \cmark  & \cmark & \cmark & \xmark \\ \hline
\multirow{1}*{\cite{zou2023wireless}} & 2023 & Integration of multi-agent GAI in wireless networks based on cloud LLMs. & \cmark  & \cmark  & \cmark  & \smark & \xmark \\ \hline
\multirow{1}*{\cite{xu2024unleashing}} & 2024 & Personalized AIGC services for mobile edge networks' privacy measures.  &  \smark  & \cmark  & \cmark & \cmark & \xmark \\ \hline

\multirow{1}*{\cite{liu2023deep}} & 2023 & Improves the efficiency of wireless network management by DGMs.  &  \xmark  & \cmark  & \cmark & \xmark & \xmark \\ \hline
\multirow{1}*{\textcolor{black}{\cite{karapantelakis2024generative}}} & \textcolor{black}{2024} & \textcolor{black}{GAI use cases across various industry mobile networks} &   \textcolor{black}{\smark}  & \textcolor{black}{\cmark}  & \textcolor{black}{\smark} & \textcolor{black}{\xmark} & \textcolor{black}{\xmark} \\ \hline
\multirow{1}*{\textcolor{black}{\cite{celik2024dawn}}} & \textcolor{black}{2024} & \textcolor{black}{Data scarcity issues in 6G Wireless through GAI}  & \textcolor{black}{\smark} & \textcolor{black}{\cmark}  & \textcolor{black}{\cmark} & \textcolor{black}{\xmark} & \textcolor{black}{\xmark}  \\ \hline
\multirow{1}*{\textcolor{black}{\cite{liu2024deep}}} & \textcolor{black}{2024} & \textcolor{black}{Enhanced physical-layer security of communication networks using GAI} &   \textcolor{black}{\xmark}  & \textcolor{black}{\cmark}  & \textcolor{black}{\xmark} & \textcolor{black}{\cmark} & \textcolor{black}{\cmark} \\ \hline
\multirow{1}*{\cite{Xia2023Aug}} & 2023 & Architecture, challenges, and outlook of GAIs. &  \cmark  & \cmark  & \xmark & \xmark & \cmark \\ \hline
\multirow{1}*{\cite{zhang2019generative}} & 2019 & GAN for autonomous driving, data completion, and traffic anomaly detection. &   \cmark  & \cmark  & \xmark & \xmark & \xmark \\ \hline
\multirow{1}*{\cite{liu2021adversarial}} & 2021 & Adversarial ML in autonomous driving, data generation, and anomaly detection. &   \smark  & \cmark  & \smark  & \cmark & \xmark\\ \hline
\multirow{1}*{\cite{cai2021generative}} & 2021 & GAN for privacy and security contexts, its benefits and challenges. &   \cmark  & \cmark  & \xmark & \cmark & \xmark \\ \hline
\multirow{1}*{\cite{deldjoo2021survey}} & 2021 & Recommender systems' security \& GANs' applications for distributed learning.  & \cmark  & \cmark  & \smark & \cmark & \xmark \\ \hline
\multirow{1}*{\cite{dunmore2023comprehensive}} & 2023 &  GANs in IDS and their applications in cybersecurity research. &   \smark  & \cmark  & \xmark & \cmark & \xmark \\ \hline 
\multirow{1}*{\cite{lin2023generative}} & 2023 & Trajectory prediction, security, and anomaly detection in transportation sectors. &  \cmark  & \cmark  & \smark & \cmark & \xmark \\ \hline

\multirow{1}*{\textcolor{black}{\cite{zhao2024generative}}} & \textcolor{black}{2024} & \textcolor{black}{GAN to enhance wireless network management}   & \textcolor{black}{\xmark} & \textcolor{black}{\cmark}  & \textcolor{black}{\cmark} & \textcolor{black}{\xmark} & \textcolor{black}{\xmark}  \\ \hline
\multirow{1}*{\cite{grassucci2023generative}} & 2023 & Generative diffusion-guided framework for SemCom, using diffusion models. &   \xmark  & \cmark  & \xmark & \cmark & \cmark \\ \hline
\multirow{1}*{\cite{thomas2023causal}} & 2023 & Causal SemCom for digital twin-based wireless systems. &   \smark  & \cmark  & \smark & \xmark & \cmark \\ \hline
\multirow{1}*{\cite{huang2023federated}} & 2023 & Enhanced privacy of AIGC using stable diffusion models and FL solutions.   & \cmark & \smark  & \cmark & \cmark & \xmark  \\ \hline
\multirow{1}*{\cite{du2023generative}} & 2023 & AaaS architecture that deploys AIGC models in edge networks. &   \xmark  & \cmark  & \cmark & \smark & \xmark \\ \hline

\textbf{This Work} &  -  & A review on applications of GAI for network management, security, and SemCom in mobile and wireless networking. & \cmark &  \cmark & \cmark  & \cmark & \cmark \\ \hline
\end{tabular}
\end{table*}

For example, GANs, a notable GAI technique, are reviewed in several recent works~\cite{yang2019generative, ayanoglu2022machine, yang2022generative, cao2023comprehensive, zhang2023complete, gozalo2023chatgpt}, providing preliminary discussions about the use of this GAI-specific technique in emerging networks. More recently, GAI has been reviewed for several research issues in wireless networks. For instance, in \cite{sabuhi2021applications}, the authors provide a detailed survey of training solutions for GANs that deal with challenges of model collapsing and gradient vanishing, providing meaningful techniques and guidance for mobile network deployments. Similarly, in~\cite{jabbar2021survey}, the authors analyze GANs with various stability issues and survey several training solutions by assessing the obstacles.
In \cite{navidan2021generative}, several critical GAN analyses of computer and communication networks are outlined, with a focus on network management, security, and mobile networks. Differently, the authors in~\cite{zou2023wireless} introduce the concept of on-device GAI, where multiple devices, functioning as distinct learning agents, deploy on-device GAI models for the collaborative processing of complex tasks.  In \cite{xu2024unleashing}, a direction on GAI services at the network edge has been outlined in various use cases, followed by guidance on building a hierarchical collaboration architecture of IoT devices, mobile edges, and centralized clouds. 

In \cite{liu2023deep}, wireless network management efficiency can be improved by DL models in general and particular applications of DGM. In addition, the authors in~\cite{van2024generative} present a review of the applications of GAI for physical layer communications, focusing on emerging problems such as signal detection, physical layer security, and Joint Source-Channel Coding (JSCC). Considering a different perspective, the authors in~\cite{liu2024generative} discuss the potential of common GAI techniques, e.g., GANs, Variational Autoencoders (VAEs), and normalizing flows, in addressing the design challenges of unmanned vehicle swarms.
{\color{black}In \cite{karapantelakis2024generative}, the authors overview the classification of GAI solutions used in mobile networks and outline the involved performance requirements. 
The authors in \cite{celik2024dawn} discuss the use cases of discriminative AI in Sixth-Generation (6G) applications and services. They then detail how GAI enhances or complements discriminative AI models across various domains, including physical layer design, network management, cross-layer network security, and localization and positioning. Based on these analyses, they conclude by emphasizing GAI's core roles in cutting-edge areas such as massive Multiple-Input Multiple-Output (MIMO), THz communications, integrated sensing and communications, near-field communications, digital twins, SemCom, AI-generated content services, mobile edge computing and edge-AI, adversarial ML, and trustworthy AI. In \cite{liu2024deep}, the authors provide a tutorial and case study of using deep generation models for wireless network management, followed by guidelines on how GAI overcomes major issues of training data scarcity, limited flexibility, and dynamic network states.}

On another front, the interplay of GAI and SemCom has opened up a new avenue for reducing resource consumption while maintaining the reliability of wireless communication networks \cite{Xia2023Aug}. SemCom relies on two fundamental techniques \cite{Yang2023Nov, Lan2021Dec} of knowledge extraction and lossy data compression, both rooted in AI. The current landscape of SemCom designs represents a paradigm shift, with these innovative concepts emerging across various disciplines. In the literature, several surveys aim to elucidate the intricate connections between the potential AI techniques within GAI and their application in the field of SemCom. From the perspective of Knowledge Graph (KG), the current SemCom is based on the KG embeddings from Natural Language Processing (NLP), which are comprehensively reviewed in \cite{Wang2017Sep, Ji2021Apr, Guo2020Oct, Guan2022Jun}. From the perspective of lossy compression, no studies have surveyed these techniques in general and lossy compression-aided SemCom in particular, despite a huge range of research that has both integrated and without knowledge representation to improve data reconstruction performance.   

In addition to the aforementioned dominant features of GAI, it also possesses significant merits in the cutting-edge area of security. For instance, in~\cite{zhang2019generative}, Zhang et \textit{al.} explore GANs in transportation, including autonomous driving, data completion, and traffic anomaly detection. In \cite{liu2021adversarial}, Liu et \textit{al.} explore how adversarial ML can be both a threat and a defense mechanism in mobile networks. In \cite{cai2021generative}, Cai et \textit{al.} provide a discussion of the potential benefits and challenges of GANs in the context of privacy and security in mobile networks. In \cite{deldjoo2021survey}, Deldjoo et \textit{al.} present the role and application of GANs in improving data distribution learning of recommendation systems. In \cite{dunmore2023comprehensive}, Dunmore et \textit{al.} survey and discuss the potential of GAN applications in enhancing cybersecurity in mobile networks, particularly in Intrusion Detection Systems (IDS). In~\cite{lin2023generative}, Lin et \textit{al.} investigate and inspect the use of GANs in enhancing trajectory prediction, security, and anomaly detection of the wireless-based transportation domain. 
{\color{black}Similarly, Zhao et \textit{al.} in \cite{zhao2024generative} also emphasize the role of advanced GAI approaches in dynamically enhancing the security, authentication, availability, resilience, and integrity of communications in the physical layer of communication networks where traditional AI approaches often struggle with the evolving physical properties of the transmission channels and emerging cyber threats.} 
However, a recent study in \cite{du2023spear} points out that the use of GAI exposes potential security vulnerabilities, and this will generate a double-edged sword for wireless security if GAI is not exploited properly.

\subsection{Motivations and Contributions}
{\color{black}It can be observed that existing surveys mostly focus on specific applications of GAI on the challenges of integrating GAI into mobile and wireless networking. However, current studies do not provide comprehensive evaluation, roles, and applications of GAI in mobile and wireless networking (i.e., GAI for networks).} To the best of the authors' knowledge, the fundamentals and applications of GAI as outlined above have not been thoroughly examined in the literature. Consequently, we aim to bridge this gap by providing the necessary preliminaries to GAI and its key applications in the future of IoT and mobile networking. This work centralizes all recent developments in GAI and complementary technologies and remarkable advancements in various aspects of mobile networks. Moreover, this work also seeks to gather the recent trending research in the field and provide a unified repository. The notable contributions of our work are listed below.
\begin{itemize}
    \item We present an overview of GAI, covering its historical development, foundational structures (e.g., autoencoders, GANs, and diffusion models), data creation, and multimodal aspects (e.g., meta-learning and multi-tasking). 
    
    \item We summarize the contributions of GAI to network management, particularly in anomaly detection and security through its ability to learn and flag deviations. 
    
    \item We explore the applications of GAI in wireless security based on baseline models, detecting anomalies, fortifying against cyber threats, improving intrusion detection, and optimizing security policies.
    
    \item We examine the applications of GAI in enhancing SemCom through NLP for human-like text, chatbots, virtual assistants, automated content creation, and improved language translation.
    
    \item We reveal challenges of GAI in mobile networking, emphasizing technical complexities, legacy systems, ethical concerns, security, and privacy and necessitating comprehensive solutions.
\end{itemize}

\begin{figure}[!t]
    \centering
    \includegraphics[width=0.98\linewidth]{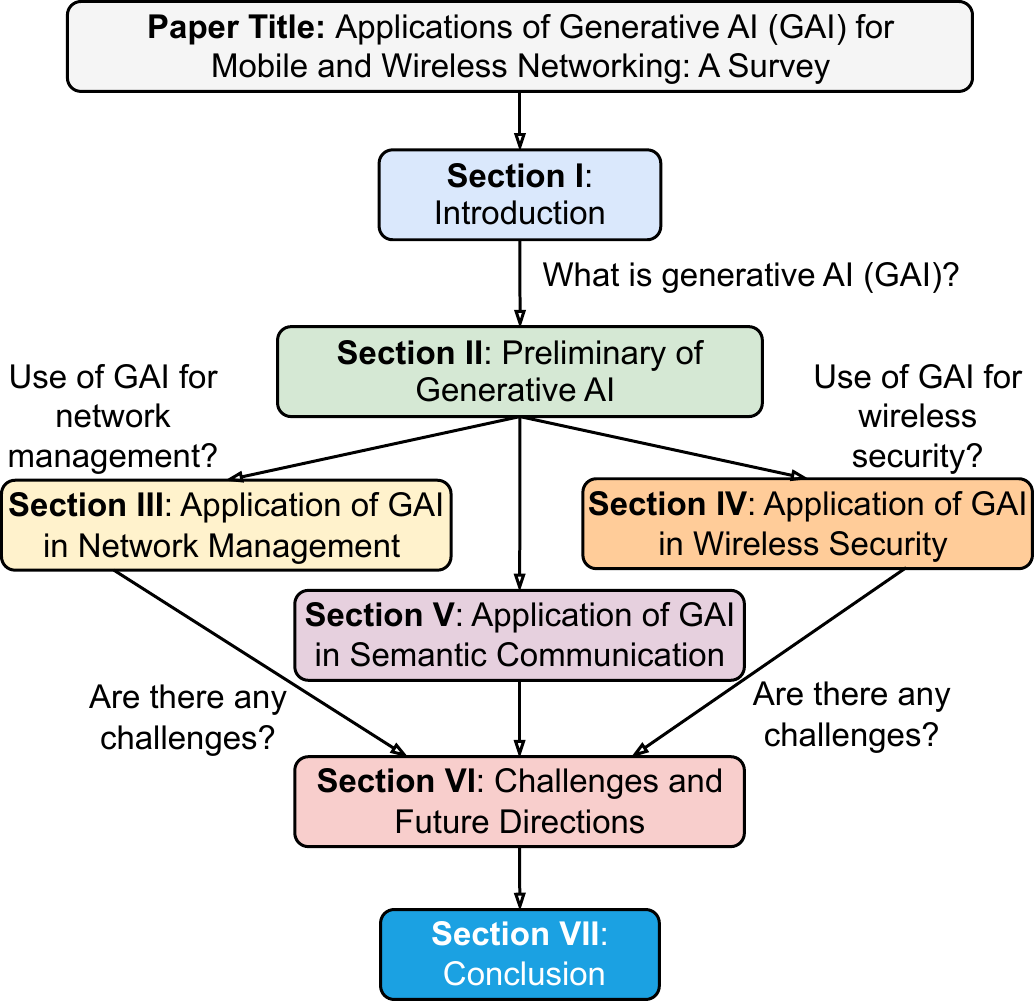}
    \caption{\textcolor{black}{The outline of this survey paper.}}
    \label{figure_structure}
\end{figure}

\subsection{Structure of the Survey}
{\color{black}The structure of this survey is summarized in Fig.~\ref{figure_structure}. Specifically, we start with a preliminary of GAIs in Section~\ref{Sec:Preliminary}, covering the development history, data-processing and data-creation techniques, and representative models. Then, we delve into the role details of GAI in mobile network management (Section~\ref{Sec:NetworkManagement}), with emphasis on levels of network control, resource allocation, network routing, and channel estimations. Next, we discuss the need for GAIs to enhance wireless security networks (Section~\ref{Sec:WirelessSecurity}) in terms of data obscurity, intrusion detection, jamming attacks, and generative steganography. Moreover, we present the promising applications of GAIs in a new communication era of mobile networks and SemCom in Section~\ref{Sec:SemCom}. Finally, we summarize the challenges and potential research directions of integrating GAIs into mobile networking in  Section~\ref{Sec:challenges}, and we conclude the paper in  Section~\ref{sec:Conclusion}.}

\section{Preliminary of Generative AI} 
\label{Sec:Preliminary}
{\color{black}
This section presents a preliminary of AI, including its evolutionary history, primitive structure, multi-modal functionality, and representative models. It serves as a basic guide for general readers to understand the following sections of GAI applications. For ease of understanding, Fig.~\ref{Figure_evolution} summarizes all information related to GAI.
}

\subsection{History of Generative AI}
{\color{black}Generative AI is rooted in ML and DL algorithms, which have evolved since the 1950s. The term ``\textit{machine learning}'' was coined by Arthur Samuel in 1952 when he developed an algorithm for playing checkers. In 1957, Frank Rosenblatt, a psychologist at Cornell University, invented the first ``\textit{neural network}'' capable of being trained, known as the Perceptron, which had a single layer with adjustable thresholds and weights. A decade later, in 1961, Joseph Weizenbaum created ELIZA, an early prototype of GAI and one of the first chatbots. ELIZA was designed to simulate empathetic conversation using natural language processing. During the 1960s and '70s, Ann B. Lesk, Leon D. Harmon, and A. J. Goldstein made significant strides in computer vision (CV) and pattern recognition by identifying 21 specific human signs. In the 1970s, Seppo Linnainmaa introduced the term ``\textit{backpropagation}'', a technique involving three main steps: processing at the output end, distributing backward, and moving through the network’s layers for training and learning. This method laid the groundwork for training Deep Neural Networks (DNNs). 

The distinction between ML and AI became evident during two ``\textit{AI winter}'' periods. The first AI winter, from roughly 1973 to 1979, saw little progress despite high expectations. However, ML continued to advance, proving cost-effective and useful for business applications like automated phone systems. In 1979, Kunihiko Fukushima developed the first multilayered Artificial Neural Network (ANN), enabling computers to recognize visual patterns and handwritten characters. The second AI winter, from around 1984 to 1990, was marked by slow development and widespread skepticism, leading to reduced funding for AI and DL research. In the late 1980s, the integration of Metal Oxide Semiconductors and Very Large Scale Integration technology resulted in efficient ANNs. In 1989, Yann LeCun's neural networks for recognizing handwritten ZIP codes marked a significant breakthrough. This progress continued with the presence of long short-term memory (LSTM) networks in 1997 and advanced GPUs by Nvidia in 1999. These advancements paved the way for launching the first digital virtual assistant, Siri, on October 4, 2011. 

The use of chatbots spiked, along with the introduction of  VAEs in 2013 \cite{kingma2013auto} and GANs in 2014 \cite{goodfellow2014generative}, both of which were built on top of DNN architectures capable of learning generative models for complex data types such as images. Upon this progress, in 2017, Google introduced the ``\textit{Transformer}'' architecture~\cite{vaswani2017attention} to enhance the robustness of GAI in learning generative models, followed by the first Generative Pre-trained Transformer (GPT), as known as GPT-1, in 2018.
Four years later, OpenAI introduced ChatGPT, a GAI model leveraging LLM, achieving new levels of performance in research, writing, and content creation. Since then, the use of GAI has spread from academia to industry, with diverse applications including infotainment, finance, advertising, defence, agriculture, and telecommunications. However, everything has two sides, and so does GAI. Table~\ref{tab:gai} covers the advantages and challenges of GAI use cases. To better understand the core value of GAI, we next go into the fundamentals of GAIs.}

\begin{table*}[ht!]
\centering
\caption{Advantages and challenges of GAI use cases based on its characteristics.}
\label{tab:gai}
{\renewcommand\arraystretch{1}\begin{tabular}{|p{0.7cm}|p{2.2cm}|p{3cm}|p{5cm}|p{5cm}|}
\hline
\textbf{Ref.} & 
\multicolumn{1}{c|}{\textbf{Attributes}} & 
\multicolumn{1}{c|}{\textbf{Description}}   & 
\multicolumn{1}{c|}{\textbf{Advantages}} & 
\multicolumn{1}{c|}{\textbf{Challenges}} \\ \hline

\multirow{4}*{\cite{pyrkov2023quantum} }&  \multirow{4}*{Noisy data }& 
Handles uncertainty in data generation to maintain the stability of results.  & 

    \textbullet~Generated diverse and variable data.  \par
    \textbullet~Recovered data during training.
    & 

        \textbullet~Inconsistent output quality. \par
       \textbullet~Unexpected results.
        
        \\ \hline
\multirow{4}*{\cite{li2020activitygan} }& \multirow{4}*{Data augmentation }& 
Improves training datasets by generating additional data samples to better train the model. & 
     
         \textbullet~Increased size and diversity of the training dataset. \par
    \textbullet~Overfitting reduction.
    & 
    
        \textbullet~High computational resources and time for generating augmented data.   \par
    \textbullet~Instability in model improvement.
    \\ \hline
\multirow{4}*{\cite{soleimani2021cross} }& \multirow{4}*{Transfer learning }       & 
Leverages pre-trained models to generate new data for specific tasks without full retraining.  & 
    
         \textbullet~Improved time and resources by reusing learned representations.  \par
    \textbullet~Useful in fine-tuning models for new tasks efficiently.
    & 
    
        \textbullet~Limited to tasks similar to the pre-trained model and entirely different domains.\par
       \textbullet~Faces with domain shift or bias issues.     
        \\ \hline
\multirow{4}*{\cite{yu2020ugan}} & \multirow{4}*{Style transfer}           & 
Applies the style of one dataset to another, e.g., changing the artistic style of images.      & 
     
    \textbullet~Facilitates artistic expression by transforming content while preserving style. \par
     \textbullet~Useful in image and video editing and creative applications.
        & 
    
   \textbullet~Artifacts or distortions appear during style transfer.\par
    \textbullet~Preserving content details can be challenging for complex styles.
    \\ \hline
\multirow{4}*{\cite{zhang2023text}} & \multirow{4}*{Text-to-image}  & 
Generates images from textual descriptions, enabling AI-powered content creation.              &   

         \textbullet~Opens possibilities for generating images based on textual ideas or concepts. \par
          \textbullet~Aids in creating visual content from text for various applications.
         & 
    
        \textbullet~Difficulty in generating highly detailed images from textual descriptions.\par
        \textbullet~Limited by the textual context and language-understanding capabilities.
        \\ \hline
\multirow{4}*{\cite{angarano2023generative} }& \multirow{4}*{Super resolution  }       & 
Enhances image resolution, useful in improving image quality for various applications.         &
    
        \textbullet~Enhances image quality for better visual perception and recognition.
        
        \textbullet~Useful in medical imaging, surveillance, and content restoration.
        & 
   
           \textbullet~Increased computational complexity and resource requirements. \par
       \textbullet~May not always achieve satisfactory results for all types of images.
        \\ \hline
\multirow{4}*{\cite{Marchetti2021Apr}} &\multirow{4}*{Music generation }        & 
Creates music compositions, often employing recurrent neural networks for sequence modeling. &
   
         \textbullet~Facilitates music composition and creativity in the absence of human composers. \par
   \textbullet~Provides a wide range of music styles and genres.
        & 
   
         \textbullet~Difficulty in generating music with the same level of emotional expression as human composers. \par
        \textbullet~May produce music that lacks coherence or structure. 
        \\ \hline
\multirow{4}*{\cite{wang2023leo}} & \multirow{4}*{Video synthesis }         & 
Generates videos by predicting frames sequentially, enabling deepfake applications.    & 
    
        \textbullet~Enables video synthesis for various utilities, including entertainment and training. \par
        
        \textbullet~Allows video generation based on textual descriptions.
        & 
             
        \textbullet~Raises ethical concerns regarding fake videos or misinformation. \par        
        \textbullet~High computational and data requirements.        
        \\ \hline
\multirow{4}*{\cite{Esmaeili2023Feb}} & \multirow{4}*{Anomaly detection}        & 
Identifies anomalies in data, useful for detecting outliers or unusual patterns.             &
          
        \textbullet~Enhances the detection of rare and critical events in large datasets. \par        
        \textbullet~Useful for security and fault detection in various industries.         
        & 
    
       \textbullet~Challenges in labeling anomalies for training, leading to imbalanced datasets.
       \par        
        \textbullet~Risk of false positives and negatives based on the anomaly definition.
        \\ \hline
\multirow{4}*{\cite{gupta2023chatgpt}} & \multirow{4}*{Data privacy}             & 
This raises concerns about potential misuse, as GAI can create realistic fake content.              &

        \textbullet~Raises awareness about data privacy and the need for robust detection of fake content. \par        
        \textbullet~Encourages the development of countermeasures against misuse.         
        & 
   
      \textbullet~Potential for generating deepfake content that can be used maliciously. \par        
        \textbullet~Challenges in distinguishing between genuine and fake content in some cases.      
      \\ \hline
\end{tabular} }
\end{table*}

\subsection{Primitive Structure of GAI} 
{\color{black}The evolution of GAI spans a long history as mentioned earlier, and its recent variants have also become diverse for different applications. However, a basic architectural block of GAI includes three fundamental components: VAE, diffusion model, and GAN. The advantages and limitations of VAEs, diffusion models, and GANs are summarized in Table~\ref{tab:advlim}.}

\subsubsection{Variational Autoencoders}
\begin{figure*}[t]
    \centering
    \includegraphics[width=\linewidth]{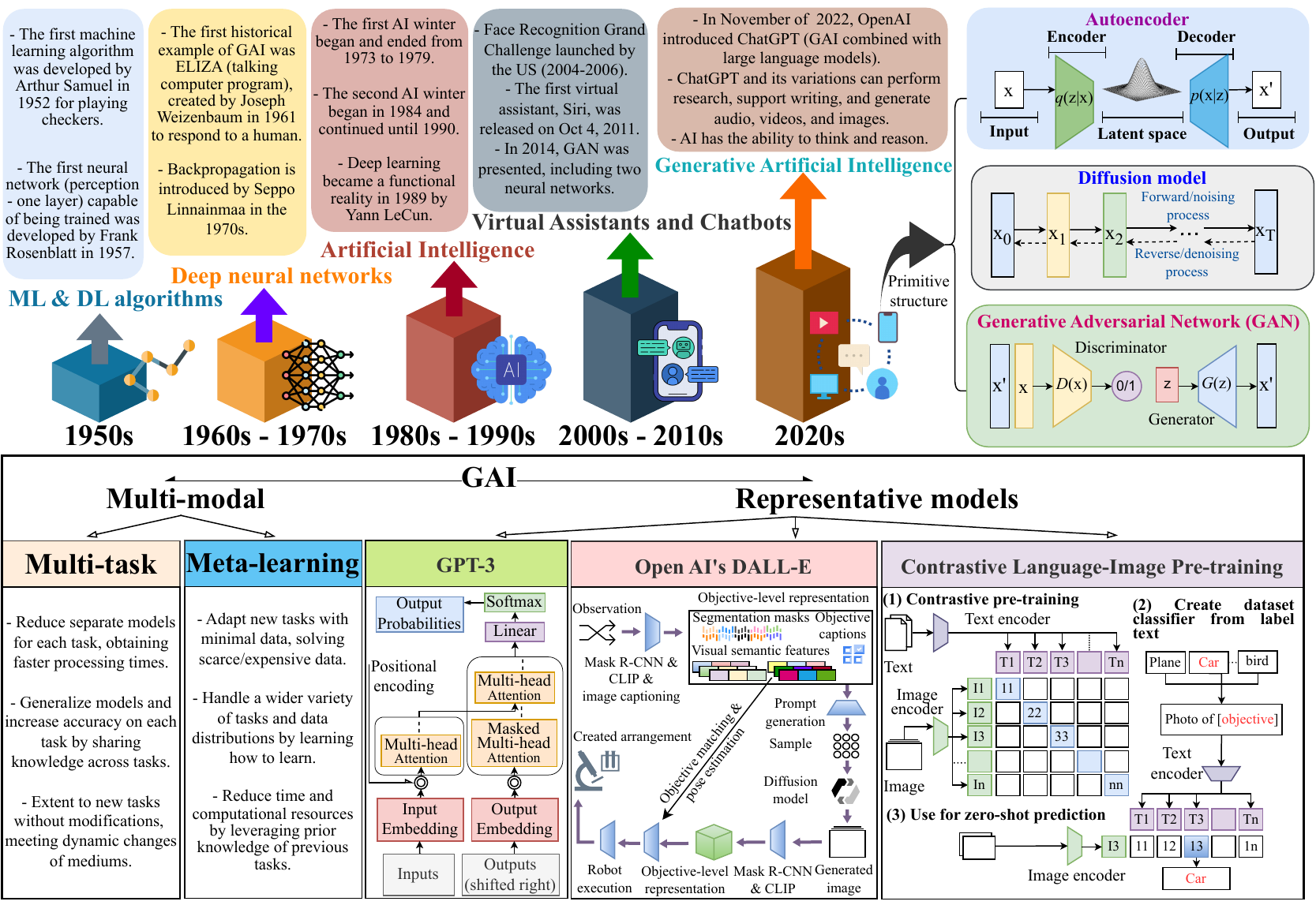}
    \caption{\textcolor{black}{Summary of GAI-involved information: The history of GAI, its primitive structure, variation of GAI, and representative GAI models.}}
    \label{Figure_evolution}
\end{figure*}

{\color{black}VAE, a family of probabilistic graphical models and variational Bayesian methods, is an ANN introduced by Diederik P. Kingma and Max Welling \cite{kingma2013auto}, as shown in Fig.~\ref{Figure_evolution}.} In VAEs' architecture, the encoder compresses data into a lower-dimensional representation, namely latent space. At the same time, the decoder module reconstructs the data from the latent space, restoring it to its original form. 
Once trained, the decoder can generate new data points by sampling from the latent space~\cite{bai2023hvae}. In GAI frameworks, VAEs are in charge of generating realistic and high-quality data, making them extremely valuable in situations with limited labeled data due to their ability to crease synthetic data samples for training purposes~\cite{ohno2020auto}. Besides, VAEs contribute significantly to dimensionality reduction and feature extraction, leading to a more compact and structured representation of complex data. 

\subsubsection{Generative Adversarial Networks}
GANs were introduced to generate realistic synthetic data \cite{goodfellow2014generative}. The key rationale behind GANs lies in their unique architecture, which consists of two neural networks: the generator and the discriminator, engaged in a competitive game. The discriminator distinguishes genuine and generated data, while the generator creates synthetic data samples that closely resemble real data. This interplay drives an adversarial training process where the generator aims to deceive the discriminator and enhance the quality of its output. {\color{black}As a result, GANs can produce diverse and realistic data since their adversarial training process can capture complex data distributions and adapt to different applications. Especially, GANs also demonstrate their unique difference from VAEs and diffusion models by their enhanced ability to control and manipulate the generated outputs through techniques such as conditional GANs \cite{Casanova2021Dec,Li2021Nov}, style transfer \cite{Kwon2022,Deng2022}, and progressive training \cite{Huynh2021}.}

\subsubsection{Diffusion Models}
{\color{black}Diffusion models are a class of probabilistic generative models,} which draw inspiration from non-equilibrium thermodynamics~\cite{Sohl-Dickstein2015Jun}. {\color{black}Technically, diffusion models consist of two distinct stages. In the first phase, a series of diffusion steps is applied to incrementally introduce random noise into the input data, where the noise magnitude varies at each step. In the subsequent phase, this process is reversed: the model iteratively learns to remove the noise, progressively reconstructing the data samples from the noisy input. As a result, during inference, data are generated by gradually reconstructing them, beginning from an initial state of random white noise. Diffusion models can be classified into three primary subcategories, with the main differences lying in the way they are used to progressively denoise the data.

\paragraph{Score-based Generative Model (SGM)} SGM is proposed by \cite{2019-DM-ALD} to address the challenges of generative models (e.g., VAEs, GANs), including: 1) mode collapse, where the model fails to capture the full diversity of data distribution, 2) instability due to the adversarial training procedure, and 3) GAN objective function is not suitable for evaluating and comparing among baselines. Besides, SGM also guides a new principle for generative modeling based on estimating and sampling from the (Stein) score \cite{2016-DM-GoFT} of the gradient of the log-density function at the input data point, i.e., vector field pointing in the direction where the log data density grows the most. The rationale behind this approach is to leverage the Langevin dynamics to move from a random sample $x_0$ to samples $x_N$ in regions of high density $p(x)$ \cite{2019-DM-ALD}. 

\paragraph{Denoising Diffusion Probabilistic Models (DDPM)} 
Neural network $\theta$ approximates the diffusion steps by leveraging $p_\theta(x_{t-1}\vert x_{t}) = \mathcal{N}(x_{t-1};\mu_\theta(x_{t},t),\Sigma_\theta(x_t,t))$ to progressively denoise the data $x_{t}$ at time step $t$, with a focus on learning to predict the noise trajectory with mean $\mu_\theta(x_{t},t)$ and covariance $\Sigma_\theta(x_t,t))$. In ideal cases, the model $\theta$ can be trained with a maximum likelihood objective so that the model inference $p_\theta(x_0)$ is as identical to each training example $x_0$. However, having $p_\theta(x_0)$ is quite intricate due to the required marginalization of all possible reverse trajectories. This promotes the study on leveraging VAE to minimize the variational lower-bound of the negative log-likelihood instead \cite{Sohl-Dickstein2015Jun}, fixing $\Sigma_\theta(x_t,t))$ to a constant value combined with rewriting $\mu_\theta(x_{t},t)$ as a function of noise  \cite{2020-DM-DDPM}, and reformulating the diffusion process as a non-Markovian process \cite{2021-DM-DDIM}. Compared to \cite{Sohl-Dickstein2015Jun} and \cite{2020-DM-DDPM}, the DDIM model in \cite{2021-DM-DDIM} achieves faster data generation with minimal compromise in data quality, for its ability to express the denoising procedure as an implicit function instead of a stochastic process. 

\paragraph{Stochastic Differential Equations} Similar to the previous two methods, the approach presented in \cite{2021-DM-SBSDE} gradually transforms the data distribution $p(x_0)$ into noise. However, it extends the previous methods by considering the diffusion process as continuous, which is modeled as the solution to a Stochastic Differential Equation (SDE). In particular, \cite{2021-DM-SBSDE} uses a neural network $\theta$ to estimate the score functions, similar to \cite{2019-DM-ALD}, and generates samples from $p(x_0)$ using SDE solvers.
}


\begin{table*}[!th]
\centering
\caption{Advantages and Limitations of GAI Models in Mobile Networking Applications.}
\label{tab:advlim}
\begin{tabular}{|p{1.5cm}|p{7.6cm}|p{7.6cm}|}
\hline
\textbf{GAI Model} & \textbf{Advantages} & \textbf{Limitations} \\ \hline
\multirow{7}*{Autoencoder}                            
& 
        \textbullet~Efficient data compression for mobile devices with limited storage.\par        
        \textbullet~Low-latency encoding and decoding suitable for real-time mobile applications.\par        
        \textbullet~Privacy preservation through on-device data processing.\par        
        \textbullet~Unsupervised learning without labeled data.\par        
        \textbullet~Well-suited for image reconstruction tasks.          
&   
        \textbullet~Limited capacity to handle diverse and dynamic mobile data.\par        
        \textbullet~Sensitivity to variations in input data quality, affecting reconstruction.\par        
        \textbullet~High computational load during training on resource-constrained mobile devices.\par        
        \textbullet~Limited diversity in generated samples.\par        
        \textbullet~Sensitive to input noise and may suffer from overfitting. 
\\ \hline
\multirow{7}*{\begin{tabular}[c]{@{}c@{}} Diffusion \\ Models\end{tabular}} 
&  
        \textbullet~Effective in modeling sequential dependencies in mobile data.\par        
        \textbullet~Privacy-preserving data generation through controlled diffusion.\par        
        \textbullet~Robust handling of diverse mobile data types.\par        
        \textbullet~Strong theoretical foundation with tractable likelihood.\par        
        \textbullet~Stable training without mode collapse.
&  
        \textbullet~Computationally demanding, challenging for mobile deployment.\par        
        \textbullet~Potential issues in capturing long-range dependencies in sequential data.\par        
        \textbullet~Training complexity increases with the scale of mobile datasets.\par        
        \textbullet~Sequential generation can be computationally expensive.\par        
        \textbullet~May struggle with capturing long-range dependencies and handling discrete data. 
        \\ \hline
\multirow{7}*{\begin{tabular}[c]{@{}c@{}}Generative\\ Adversarial \\Networks\end{tabular}} 
&      
        \textbullet~Creation of realistic synthetic data for diverse mobile applications.\par        
        \textbullet~Enhanced mobile security through adversarial training against attacks.\par        
        \textbullet~Suitable for mobile gaming and augmented reality applications.\par        
        \textbullet~High-quality sample generation in various domains.\par        
        \textbullet~Effective in learning complex data distributions.
&
        \textbullet~Training instability might lead to mode collapse.\par        
        \textbullet~Prone to generating biased data, impacting fairness in mobile applications.\par        
        \textbullet~Resource-intensive training and generation process on mobile devices.\par        
        \textbullet~Lack of explicit likelihood estimation.\par        
        \textbullet~Sensitivity to hyperparameter tuning.
        \\ \hline
\end{tabular}
\end{table*}

\subsection{Multi-modal GAI}
{\color{black} Multi-modal GAI is an advanced level of GAI that allows one to handle various tasks with different requirements. This is achieved by integrating diverse modalities like visual, auditory, and textual information \cite{suzuki2022survey}. However, due to the different uses of real-world scenarios, multimodal GAI can be differentiated into meta-learning GAI and multi-task GAI. The following discussion will explore how each method addresses specific scenarios with its methods and use cases.}

\subsubsection{Meta-learning GAI}
{\color{black}Meta-learning GAIs refer to ``\textit{learning to learn}'' models that aim to quickly adapt to new tasks with minimal data constraints by leveraging prior knowledge. This is achieved by training a model on various combinations of tasks (e.g., text, image, and audio) to learn a meta-knowledge or strategy that can be applied to new or unseen tasks. This feature becomes particularly useful in few-short or zero-short learning scenarios. 

For example, GAI systems with meta-learning can achieve fast adaptation by leveraging the task-specific gradient trajectories \cite{Finn2017Aug}.} This approach enables GAI systems to acquire knowledge more efficiently and transfer it across diverse domains \cite{Shi2021Apr}, while also possessing the capability to continue learning \cite{cao2022meta, zhao2022boosting}. Additionally, meta-learning also addresses the challenge of continuous learning by mitigating the issue of catastrophic forgetting, allowing GAI to retain previously acquired knowledge while continuously acquiring new knowledge or reducing the need for collecting a large number of real samples \cite{kim2022massive, qin2020generative}. Especially, instead of applying meta-learning to quickly adapt the GAI model from different domains, a different way of leveraging meta-learning, called MetaMask, is proposed in \cite{Li2022Dec}, with the focus of leveraging meta-learning to efficiently adapt the joint GAI model via trajectory estimation from the GAI sub-model. The taxonomy of meta-learning GAIs can be summarised by \cite{Hospedales2021May}.

\begin{figure}
    \centering
    \includegraphics[width=\linewidth]{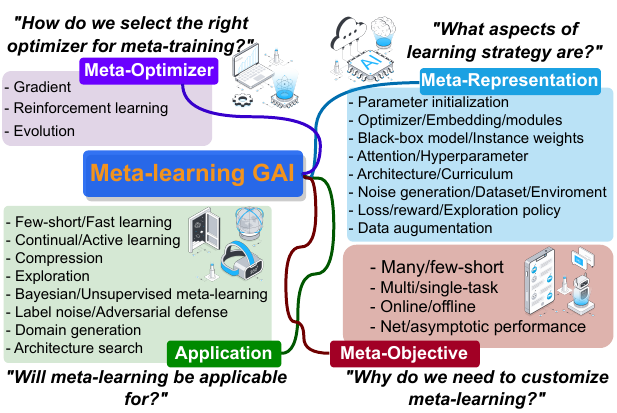}
    \caption{\color{black}Taxonomy of meta-learning through GAI.}
    \label{fig:enter-label}
\end{figure}

\subsubsection{Multi-task GAI}
Multi-task GAI refers to the ability of an AI system to effectively perform and generalize across a wide range of tasks \cite{Zhang2021Mar, Wang2020Apr, Hang2020Jun}. It utilizes transfer learning to improve performance across various domains by leveraging the knowledge and skills gained from one task and applying them to related or similar tasks~\cite{xu2023generative, guo2023zero}. Moreover, it can extract shared representations and beneficial characteristics to improve performance on new and unexplored tasks. Collaborative learning from multiple tasks reduces training time and enhances computational efficiency, optimizing resource utilization. Another advantage of multi-task GAI is domain adaptation, enabling the system to adjust its knowledge and skills to different domains or tasks~\cite{ren2020multi}. This adaptability empowers the system to achieve high performance in unfamiliar or dynamic environments. Additionally, multi-task GAI facilitates continuous learning by gradually introducing new tasks while minimizing the forgetting of previously mastered tasks. This dynamic approach allows ongoing knowledge upgrades and long-term proficiency across diverse disciplines.

\subsection{Representative GAI Models}
Across a diverse range of tasks and domains, several representative GAI models have demonstrated impressive capabilities. For example, GPT-3, a renowned model developed by OpenAI, stands out for its exceptional language processing abilities, excelling in tasks like language generation, translation, summarization, and question-answering~\cite{roumeliotis2023chatgpt}. OpenAI's DALL-E model is a specialized form of GAI that focuses on generating images from textual descriptions, effectively transforming natural language prompts into accurate visual representations~\cite{kapelyukh2023dall}. It enables tasks such as image categorization, object detection, and text-based image synthesis. \textcolor{black}{However, GPT-3 and DALL-E require overparameterization, making them unsuitable for deployment in IoT systems.}

\textcolor{black}{Inspired by this, Contrastive Language-Image Pre-training (CLIP) \cite{2021-GAI-CLIP} and Bootstrapping Language-Image Pre-training (BLIP) \cite{2022-GAI-BLIP} have emerged as promising candidates for the future of GAI in IoT networks due to twofold: 1) Lightweight architectures, and 2) Pretrained models, allowing their parameters to be effectively transferred to a variety of applications such as conditional text to image (ControlNet) \cite{2023-DM-ControlNet}, Zero-shot image to image (Pix2Pix-Zero) \cite{Pix2PixZero}.} These GAI models exemplify the potential for developing intelligent systems that excel across various activities and domains, showcasing cutting-edge progress in the field of AI.


\begin{figure*}[t]
    \centering
    \includegraphics[width=\linewidth]{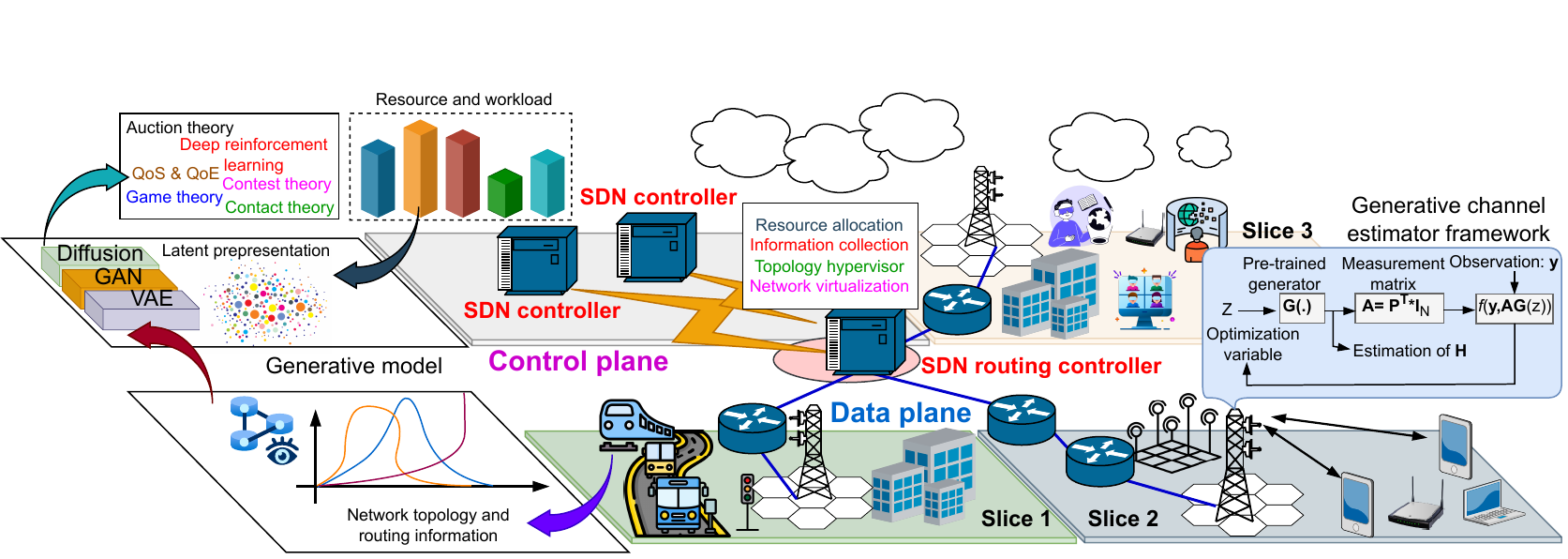}
    \caption{\textcolor{black}{Illustration of GAI applications in network management, including network control, resource allocation, and channel estimation.}}
    \label{Figure:GAI_NetMan}
\end{figure*}

\section{Application of GAI in Network Management}
\label{Sec:NetworkManagement}
{\color{black}This section reviews the applications of GAI in network management, including network control, resource allocation, and channel estimation. Fig.~\ref{Figure:GAI_NetMan} illustrates the applications of GAI in network management.  Then, we conclude this section with lessons learned and key takeaways.}

\subsection{Network Control}
{\color{black} In the context of network softwarization, network control can be classified into three levels: Software-Defined Networking (SDN), network slicing, and network routing \cite{8187644}. SDN centralizes and automates network control, allowing for dynamic adjustments and efficient resource management. Network slicing leverages SDN to create customized, isolated virtual networks tailored to specific needs, enhancing flexibility and efficiency. The centralized management of SDN facilitates network routing, a fundamental component, which ensures the best path selection for data transmissions. }

\paragraph{Software-Defined Networking}{\color{black}SDN separates the network's control plane, which makes decisions about traffic routing, from the data plane, which forwards traffic to the selected destination. Particularly, SDN introduces dominant features that enable billions of IoT-connected devices to be controlled and accessed remotely \cite{Bera2017Aug}. Thus, integrating GAI into SDN frameworks empowers network administrators with enhanced automation, intelligent resource allocation, and adaptable network optimization, thereby meeting the IoT requirements in an efficient, scalable, seamless, and cost-effective manner.} More specifically, GAI is in charge of real-time decision-making for effective traffic routing, load balancing, and Quality-of-Service (QoS) management, enabling SDN controllers to learn from and adapt to changing network conditions dynamically. For example, using a diffusion model to generate effective contracts in~\cite{liu2023deep} can address the issues of data scarcity, limited flexibility, and dynamic network states. In~\cite{chen2023netgpt}, NetGPT is proposed to employ LLMs at the edge and the cloud in conjunction with location-based information to speed up personalized procedures and cloud interaction, enhancing intelligent network management and orchestration. 

\paragraph{Network Slicing}{\color{black}Network slicing is used primarily in Fifth-Generation (5G) networks and Sixth-Generation (6G) to create multiple virtual networks on a shared physical infrastructure. Especially, in the IoT field \cite{Wijethilaka2021Mar}, network slicing can accommodate different network needs of heterogeneous IoT applications through dedicated slicing. This presents a promising research direction, where leveraging GAI helps to enhance customizing network instances within shared IoT infrastructures \cite{phyu2023machine}.} For example, GAI can exploit the virtual properties to perform network data analysis, pattern prediction, and generate optimized network slices tailored to specific requirements. In this way, wirelessly intelligent networks can dynamically allocate resources, balance workloads, and adapt to changing traffic demands, contributing to better user experiences, increased network efficiency, and greater flexibility in service provisioning~\cite{abdellatif2023intelligent}.

\paragraph{Network Routing}{\color{black} Network routing typically consumes significant bandwidth and resources, particularly when selecting optimal paths to ensure efficient data transmissions from the source to the destination in large-scale IoT environments, where frequent updates are necessary.} Introducing GAI into network routing can enhance the efficacy of choosing and optimizing routing algorithms for certain network objectives by simulating, creating, and analyzing synthetic network scenarios. An example of this is exploiting a one-shot conditional generative routing model to perform one-shot routing to the pins within each network, and the order in which the networks need to be routed is learned adaptively \cite{cheng2022policy}. Another example of GAI in routing solutions for different network status distributions and topology is exploiting a transfer RL algorithm to improve training efficiency by rapidly transferring knowledge~\cite{dong2021generative}. 

\subsection{Resource Allocation}
{\color{black}Resource allocation involves distributing available resources such as bandwidth, power, and time slots to optimize network performance for various users and services.} In the context of dynamic wireless IoT networks where access entities frequently arrive and depart in short periods and changes in wireless mediums are unpredictable, GAI contributes to dynamically changing network resources in real-time through continuous monitoring and analysis of network traffic. This ensures optimal performance and responsiveness to the evolving needs of various network factors such as network load, user demands, and QoS demands. It also helps systems use proper resource allocation strategies to ensure an optimal allocation that aligns with network objectives and user expectations. 

For instance, the authors in~\cite{du2023yolo} propose GAI-based resource allocation for the SemCom framework based on YOLOv7-X. This method helps reduce costs and improve information transfer in communication services by extracting the necessary semantic information from edge device images before deciding on resource allocation via confidence and AI-generated models. In~\cite{du2023generative}, an AIGC-as-a-Service (AaaS) architecture is introduced to wireless edge networks, where GAI  at edge servers is responsible for repairing damaged images and naturally creating high-resolution/augmented reality/virtual video content, reducing computational resources. Compared to DRLs, this architecture has not only a simple transmission protocol, but also an efficient and scalable method for mobile users to access intelligent services on demand without limited resource concerns.

\subsection{Channel Estimation}
{\color{black}Channel estimation is a process used in wireless networking to determine the characteristics of a communication channel. This information is crucial for optimizing the transmission and reception of data.} Numerous benefits can be achieved by incorporating GAI into mobile networking channel estimation. First, GAI enhances the ability to detect faulty signals and provide appropriate fixes to improve channel estimation procedures. Second, GAI can streamline the resources used for channel estimation, which lowers the computational complexity and energy usage of conventional estimation techniques. Third, by learning the fundamental patterns and properties of wireless channels, GAI makes channel estimation more reliable and precise, particularly in difficult situations with little training data or time-varying channel circumstances.

To verify these promising traits, the work in \cite{balevi2020high} designs a high dimensional channel estimation approach based on a deep generative network combined with a compressive sensing strategy. The experiments from \cite{balevi2020high} indicate that GAI not only uses fewer channel pilots than conventional techniques but also improves the execution time of channel estimation procedures for one-bit quantized pilot measurements. Similarly, another study for wideband channel estimation \cite{balevi2021wideband} has demonstrated the efficacy of GANs in dealing with channel estimation in the conditions of limited channel pilots and low transmit SNR. Specifically, it significantly reduces the number of pilots by about 70\%, even at mmWave and THz frequencies, without affecting the estimation errors. On another front, the work in \cite{arvinte2022mimo} displays the potential of deep score-based generative models to multi-input multi-output systems, with high fidelity channel estimation for channel communication of sizes 64×256 with pilot density up to 25\%.

\subsection{Lessons Learned and Key Takeaways}
{\color{black}Several insights in various network control and optimization fields have been gained from investigating GAI applications in network management. The dynamic nature of large-scale networks and their complexity are frequently beyond the capabilities of traditional models. However, GAI can learn complex network behaviors and automate decision-making processes with the help of VAEs and GANs. As a result, GAI-based communication networks can respond more quickly in real-time to changes in traffic, load, or performance conditions.

By anticipating traffic patterns and dynamically modifying flow rules to maximize network performance, GAI facilitates the control plane within SDN to be more responsive and agile, decreasing latency and increasing total reaction times. Besides, GAI enhances the management of network slices by modeling various user needs and network circumstances. This enables more precise and effective resource allocation across various network slices, ensuring high-quality service while reducing resource waste.

GAI has also demonstrated its effectiveness in providing more accurate forecasts of network demands for resource allocation. This allows for adaptive allocation strategies that maximize network usage and improve QoS. GAI largely simplifies resource allocation procedures through reinforcement learning-based techniques, especially when network demands fluctuate. GAI’s ability to create models from sparse or insufficient data benefits channel estimation. It is therefore possible to forecast channel conditions more precisely, which enhances communication reliability—particularly in challenging situations where conventional approaches would be insufficient.}

{\color{black} The key takeaways from this section are as follows.
\begin{itemize}
    \item \textbf{Key 1.} GAI can enhance network control by employing predictive models to optimize traffic flow and minimize time (e.g., training time and execution time).

    \item \textbf{Key 2.} GAI offers flexible network routing by extracting information learned from synthetic simulation scenarios with combination of real-time monitoring of data pathways.

    \item \textbf{Key 3.} By dynamically adjusting computing resources based on real-time demand and ensuring optimal efficiency, the resource allocation capabilities are improved. 

    \item \textbf{Key 4.} Generative models improve estimation accuracy, resulting in significant benefits for channel estimation.
\end{itemize}
}

\section{Application of GAI in Wireless Security}
\label{Sec:WirelessSecurity}
\begin{figure*}[ht!]
  \centering \includegraphics[width=\linewidth]{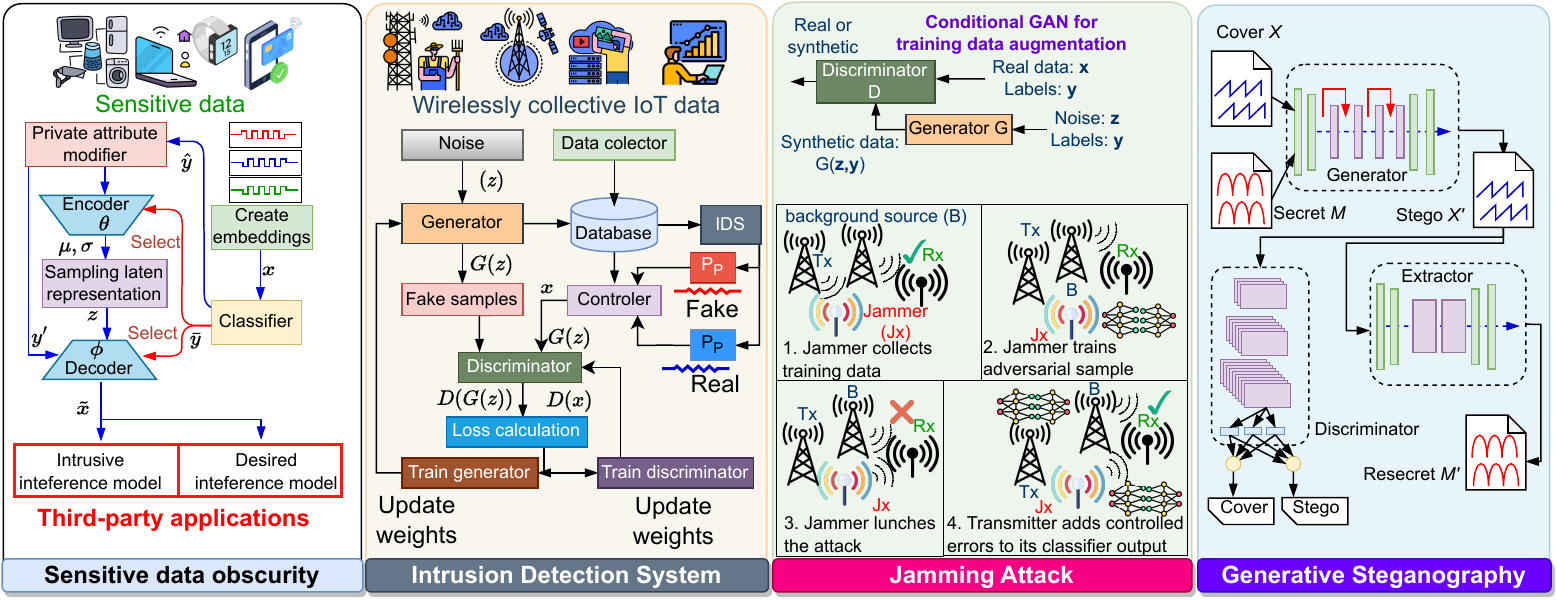}
   \caption{\color{black}Illustration of GAI applications in wireless security: sensitive data obscurity, intrusion detection system, jamming attack, and generative steganography.}
   \label{fig:security}
\end{figure*}

{\color{black}This section reviews the applications of GAI in dealing with security issues of mobile and wireless networking, covering from sensitive data obscurity, intrusion detection, and jamming attacks, to generative steganography (See Fig.~\ref{fig:security}). Then, this section concludes with useful lessons and key takeaways.}

\subsection{Sensitive Data Obscurity}
{\color{black}Sensitive data obscurity refers to techniques that hide or obscure sensitive information to protect it from unauthorized access or eavesdropping, such as encryption and data masking \cite{Aouedi2024Jul}. This ensures that data remains unreadable and secure even if it is intercepted. In the context of large-scale IoT particularly in mobile and wireless networking, sharing raw data between intelligent systems for training machine learning models without proper safeguarding strategies poses numerous risks to user privacy, where sensitive information may be disclosed by third parties through unauthorized access or interception, as seen in the case of sensors in IoT devices \cite{Hajihassnai2021May}.} Recently, GAI has demonstrated its promising capabilities in obfuscating sensitive data by creating synthetic data that closely mimics real datasets~\cite{dutta2020generative}.

There are multiple methods for generating synthetic data using GAI models. Data perturbation is one of GAI's main techniques in safely hiding sensitive data \cite{khaliq2022secure}. This technique uses synthetic data to complicate extraction of useful information, by applying controlled perturbations to sensitive data~\cite{wan2022defense}. In addition, distributing and transforming only the sensitive elements in datasets can also increase the security of data sharing while preserving the features and statistical utility of the original data for model training activities \cite{shojaee2022task}. Moreover, GAI models can apply other techniques like differential privacy and data anonymization. For example, in encrypted communication, indistinguishable noise can be created from encrypted data using generative models, complicating decryption of the original data without the proper decryption keys~\cite{ferrag2023generative}. With this strategy, sensitive data are sent more securely over wireless networks while maintaining their confidentiality and integrity.

\subsection{Intrusion Detection}
{\color{black} Intrusion detection is a process of monitoring and analyzing events in wireless-secured IoT networks to identify signs of unauthorized access or malicious activity. IDS is a valuable tool for detecting these threats by monitoring network traffic and raising system alarms when suspicious activity is identified. Due to its predominant features}, applying GAI to IDSs opens new avenues to enhance detection capabilities and improve wireless network security~\cite{park2022enhanced}. GAN is a valuable tool for discovering anomalies and distinguishing malicious activity in wireless network traffic assisted by GAI models. GANs are capable of learning the underlying patterns and behaviors of network traffic through training on large-scale IoT datasets, allowing them to detect unforeseen attack patterns from expected norms with high accuracy~\cite{seo2018gids}.

Generating synthetic data samples that closely resemble actual network traffic is one of the main benefits of utilizing GAI in intrusion detection. IDS is more resilient and able to handle different types of threats when it uses GAIs to improve its training data and create a range of testing and evaluation scenarios based on fake attack patterns. The early discovery of zero-day attacks, vulnerabilities or exploits of which security experts are unaware, can also be helped by GAI~\cite{huang2023zero}, where IDS exploits generative models to detect suspicious actions that depart from usual behavior and generates an early warning action to the system for potential zero-day attacks. Additionally, IDSs-based GAI approaches can improve their detection accuracy and minimize the impact of false alarms by lowering false positives and false negatives~\cite{yadav2021survey}. Similarly, the use of generative models can lower the computing burden of IDS, enabling instantaneous network traffic monitoring and analysis~\cite{yang2022federated}.

\subsection{Jamming Attacks}
{\color{black}Jamming attacks refer to the intentional transmission of interfering signals by attackers to disrupt or block wireless communication, hindering legitimate devices to communicate due to poor reception \cite{pirayesh2022jamming}. In the context of IoT where multiple networks coexist, exploiting GAI can help recognize and prevent jamming attacks by taking necessary actions when they occur, for its ability to learn and comprehend the underlying patterns and characteristics of wireless signals.

For example, GAI can help mitigate jamming action by making the transmitter’s behaviour unpredictable 
\cite{Erpek2018Dec}, such as taking deliberately wrong transmissions in some selected time slots  (i.e., transmitting on a busy/idle channel) in combination with the ML algorithm for spectrum sensing classification. This results in an attack in which the transmitter returns to the jammer, deceiving the adversary's surveillance capabilities.
Besides, exploiting GAI brings considerable benefits to secure cognitive radio networks by extracting multi-signal representation from a wideband spectrum with a high sampling rate and consequent high-dimensional data  \cite{toma2020ai}.  Likewise, exploiting GAI can help the receiver classify accurate timing and positioning information in the navigation task from distinguishing between spoofed and authentic satellite signals \cite{Li2021Aug}.
Another example is GAI-assisted jamming mitigation systems trained on diverse signals, which can identify and categorize various types of jamming attacks~\cite{ayanoglu2022machine}.}





\subsection{Generative Steganography}
{\color{black}Generative steganography is a data-hiding method, where secret data is directly used to generate stego media, such as images, without requiring a cover image \cite{Shi2018May}. In light of wirelessly intelligent IoT networks, where data privacy and sensitive information security are top priorities, the combination of GAI and steganography enhances safeguarding sensitive information in seemingly harmless wireless signals~\cite{liu2020recent}. In this approach, GAI uses GANs, trained on the statistical properties of typical wireless signals~\cite{Ambika2022Dec}, to generate realistic images and embed secure information directly into the pixel values, and the steganography techniques provide methods to securely embed and extract this hidden data.}

In this way, bringing GAI-combined steganography into mobile and wireless networking offers several benefits. For example, this approach provides a covert way to send sensitive data without prompting suspicion~\cite{dutta2020generative}. Besides, GAI-combined steganography improves mobile networks' resistance to unauthorized access and data breaches \cite{tan2021channel}. Even if the attackers can successfully intercept the transmission, decrypting the secret data integrated into the cover signals without proper tools is impractical. Furthermore, generative steganography offers a strong defense against assaults meant to compromise the secrecy and authenticity of wireless communications. The process of embedding and extracting concealed information can be extremely efficient and resilient to detection by unauthorized parties using GAI models~\cite{Chen2021Aug}.

\begin{table*}[ht]
\centering
\textcolor{black}{
\caption{GAI Use Cases in Wireless Network Security}
\label{tab:use}
\resizebox{\textwidth}{!}{
\begin{tabular}{|p{3.7cm}|p{1.5cm}|p{4.0cm}|p{4cm}|p{4.0cm}|}
\hline
\textbf{Security Applications}        
& \textbf{Techniques}    
& \textbf{Problems Addressed}                       
& \textbf{Observed Benefits}                                     
& \textbf{Tangible Impact}                                    
\\ \hline
Sensitive data obscurity~\cite{Hajihassnai2021May}             
& VAEs                      
& Safeguarding private information                  
& Improved anonymization of data                                  
& Enhanced adherence to privacy laws                           
\\ \hline
Intrusion detection~\cite{park2022enhanced}                 
& GANs                      
& Identifying any unapproved access                  
& Enhanced precision in identifying breaches                      
& Increased robustness of the network                        
\\ \hline
Jamming attacks~\cite{toma2020ai}                     
& Diffusion Models          
& Reducing interference caused by jammers            
& Real-time jamming attack mitigation               
& Better accessibility of the network                         
\\ \hline
Generative steganography~\cite{Lee2023Apr}             
& GANs                     
& Data hiding mechanisms                             
& Improved data protection and camouflage                          
& Routes of secure communication                              
\\ \hline
Secure data transmission~\cite{zhang2019generative}            
& Autoencoders              
& Hiding sensitive data during transmission 
& Improved security of data in transit                             
& Decreased chance of data leaks                             
\\ \hline
Traffic anomaly detection~\cite{van2024generative}   
& GANs                      
& Recognizing unusual traffic trends                 
& Enhanced identification of possible threats                      
& Enhanced security protocols for networks                    
\\ \hline
Physical layer security~\cite{liu2024deep}            
& Diffusion Models          
& Addressing the physical layer's vulnerabilities     
& Increased channel security                    
& Enhanced honesty and secrecy 
\\ \hline
\end{tabular}
}
}
\end{table*}

\begin{table*}[!th]
\centering
\caption{GAI for Optimizing Wireless Security Enforcement Scenarios.}
\label{tab:GAI_Security}
\resizebox{\textwidth}{!}{
\begin{tabular}{|p{0.5cm}|p{0.65cm}|p{1.25cm}|p{2.5cm}|p{2.8cm}|p{1.1cm}|p{3.5cm}|p{4.4cm}|}
\hline
\textbf{Apps}               
& \textbf{Ref} 
& \textbf{Model}
& \textbf{Inputs}
& \textbf{Outputs} 
& \textbf{Loss function}
& \textbf{Performance} 
& \textbf{\textcolor{black}{{Limitations}}} \\ \hline
                          
\multirow{10}{*}{\rotatebox[origin=r]{90}{Sensitive Data Obscurity}}  &  \multirow{3}*{\cite{clark2019privacy} }           
&   GAP-GAN            
&   Spectrum sharing policies              
&   Obfuscated data for adversary's inference               
&   Minimax                   
&   Balances attacks and utility for spectrum sharing and signal mapping                   & \textcolor{black}{GAN architecture may lead to mode collapse, where the privatizer could dominate the adversary.} \\ \cline{2-8} 
&  \multirow{3}*{\cite{rahman2021secure}}            
&  Federated DL              
&   Decentralized traffic statistics               
& Botnet attack detection and anomaly detection results                 
&    MSE                    
&   Outperforms standard centrally managed system in attack detection accuracy                  & \textcolor{black}{Models may suffer from data heterogeneity.} \\ \cline{2-8} 
         &  \multirow{2}*{\cite{mohanty2022robust} }           & Stacking ensemble               &  Darknet traffic data               &  Darknet traffic identification and characterization                &        MSE               &   Robust against four adversarial attacks                  & \textcolor{black}{The learners are simple; thus, not robust on complicated data.} \\ \cline{2-8} 
\multirow{16}{*}{\rotatebox[origin=r]{90}{Intrusion Detection}} & \multirow{3}*{\cite{xu2020toward}}             &  CVAE              &   Intrusion data              & Intrusion detection results                 &   log-cosh loss term                     &   Capable of generating new intrusion data with promising diversity                   & \textcolor{black}{Requirements of labels make the method infeasible when applied to real data that lacks human labellings.} \\ \hline
     &  \multirow{2}*{\cite{yang2019improving} }           &  CVAE              & Network data features                &  Intrusion detection results                & Max                       &   Better overall accuracy, detection rate, and false positive rate                  & \textcolor{black}{Requirements of label make the method infeasible when applied to real data that lacks human labellings.} \\ \cline{2-8} 
    &  \multirow{3}*{\cite{park2022enhanced}}            &    GAN            &  Heterogeneous network data               &   Intrusion detection results               &   GAN Loss                  &    Efficiently detects network threats and resolves data imbalance problems                 & \textcolor{black}{GAN architecture may lead to mode collapse, where the privatizer could dominate the adversary.} \\ \cline{2-8} 
    &  \multirow{2}*{\cite{ding2022gan}}            &  GAN      &   Metaverse network traffic data              &   IoT intrusion detection                 & GAN Loss                 & Higher accuracy for binary and multiple classification                     & \textcolor{black}{GAN architecture may lead to mode collapse where the privatizer could dominate the adversary.} \\ \cline{2-8} 
    &  \multirow{2}*{\cite{ferrag2023generative}}            &  Transformer-based GAN            &   Data from 6G-enabled IoT              &  Cyber attack detection results                &   GAN Loss           & Achieved high overall accuracy in detecting IoT attacks                     & \textcolor{black}{GAN architecture may lead to mode collapse, where the privatizer could dominate the adversary.} \\ \hline
\multirow{11}{*}{\rotatebox[origin=r]{90}{Jamming Attacks}}  &   \multirow{3}*{\cite{jayabalan2023generative} }          & GDNN, GALNN               &   Communication data from cognitive radios              &  Transmission decisions to mitigate jamming attacks                &   Cross entropy                     &   Achieves successful transmission by taking optimal actions for defense mechanism   & \textcolor{black}{GAN architecture may lead to mode collapse, where the privatizer could dominate the adversary.} \\ \cline{2-8} 
    & \multirow{2}*{\cite{han2021better}   }          &  GAN-enhanced              &  Spectrum sensing information               &  Reconstructed spectrum waterfall &     Adversarial \& reconstruction                &  Effectively avoids complex jamming attacks                   & \textcolor{black}{The need for data accumulation in the experience memory causes the system to suffer from straggling.} \\ \cline{2-8} 
     & \multirow{3}*{\cite{shi2019generative} }            &   GAN-based              &  Modulation signal &  Probability of misclassification for spoofing attacks                &  Minimax                       &    Significant increase in the success probability of wireless signal spoofing & \textcolor{black}{The model is fixed on the trained modulation signals, and may not generalize to other signals.} \\ \cline{2-8} 
    & \multirow{3}*{\cite{zhang2023detection} }            &  Hidden Markov model               &   Signals processed with sliding window              &  Jamming status of each received energy sample                & MSE &  Demonstrates superiority in detecting stealthy jamming without prior knowledge  & \textcolor{black}{The simple model makes the learning not generalized when detecting the jammers with different behaviors.} \\ \hline
\multirow{10}{*}{\rotatebox[origin=r]{90}{Generative Steganography}}     &  \multirow{2}*{\cite{Ambika2022Dec}}            &   GAN based             &   Secret information with noise vector              &  Stegno image containing hidden secret information                &      Minimax                  &  Evaluation shows improved image quality and security              & \textcolor{black}{A huge amount of data is required to train the GAN model.} \\ \cline{2-8} 
    & \multirow{2}*{\cite{wei2022generative} }            &   GSN             &  Secret and noise data                &   Realistic stego images with secret data               &   GAN loss                     &    Superior resistance to steganalysis detection                & \textcolor{black}{The model may suffer from mode collapse, where the data is overfitted into another data.} \\ \cline{2-8} 
     & \multirow{3}*{\cite{zhang2019generative} }            &  DCGAN               & Secret message, uncorrupted region in the image             &  Stego image with hidden secret message                &     Contextual \& perceptual                      &  Qualitative and quantitative evaluations of the generated stego images. & \textcolor{black}{The model may suffer from mode collapse, where the data is overfitted into another data.} \\ \cline{2-8} 
    
    & \multirow{3}*{\cite{Chen2021Aug} }            &  GAN              &   Dealer's distribution and human face images              & Revealed message when authorized participants grant consent                 &         Noise vector             & Secure message authentication and multi-factor authentication                     &  \textcolor{black}{The training process needs high computational resources, and the model is large, making it infeasible for IoT}\\ \hline
\end{tabular}
}
\end{table*}

\subsection{Lessons Learned and Key Takeaways}
Numerous applications of GAI in wireless security have demonstrated promise for bolstering mobile networking defenses against a variety of threats and weaknesses. Differential privacy and safe multi-party computing are two methods that have been successful in obscuring critical information and preserving user privacy. Network security is strengthened against DoS attacks and zero-day vulnerabilities due to the integration of GAI models into network IDS, which enables quick detection of unauthorized access attempts and security breaches. To guarantee network integrity, GAI's proactive and dynamic intrusion detection algorithms are essential.

In addition, recent developments in GAI have proven crucial in the fight against jamming attempts that interfere with wireless communication. GAI improves malware detection and stops the propagation of threats by analyzing network data and device behavior to quickly identify and isolate infected devices. Also, by encrypting data and disguising it within seemingly innocent material, steganography's novel application of GAI enables secure data transmission. These uses serve as a reminder of the importance of GAI for improving data security and facilitating safe and secure user verification.

Mobile networking offers more robust security measures due to the adaptability of GAI in varied security concerns. GAI approaches hold promise to increase the state-of-the-art in mobile network security and communication and data transmission integrity by strengthening defenses against sophisticated attacks. Various use cases of GAI approaches in wireless network security can be compiled in Table~\ref{tab:use}. It demonstrates how several GAI models solve particular security issues, improving network resilience, threat detection, and data protection. In addition, our review uncovered numerous innovative research methods that further demonstrate the potential of GAI to offer a thorough defence against changing threats and improve the mobile networking security environment, as summarized in Table~\ref{tab:GAI_Security}. The future of mobile networking security is bright due to ongoing developments in GAI, which will ensure a safer and more secure digital environment.

{\color{black}The key takeaways from this section are as follows.
\begin{itemize}
    \item \textbf{Key 1.} GAI strengthens wireless security against anomalies and intrusions in realtime.

    \item \textbf{Key 2.} GAI combats jamming attacks and ensures secure data transmission with privacy protection.

    \item \textbf{Key 3.} GAI achieves enhanced information encryption through the steganography technique.

    \item \textbf{Key 4.} Ongoing research in GAI provides more resilient security measures for mobile networking.
\end{itemize}
}

\section{Application of GAI in Semantic Communication}
\label{Sec:SemCom}

\begin{figure}[t!]
  \centering \includegraphics[width=\linewidth]{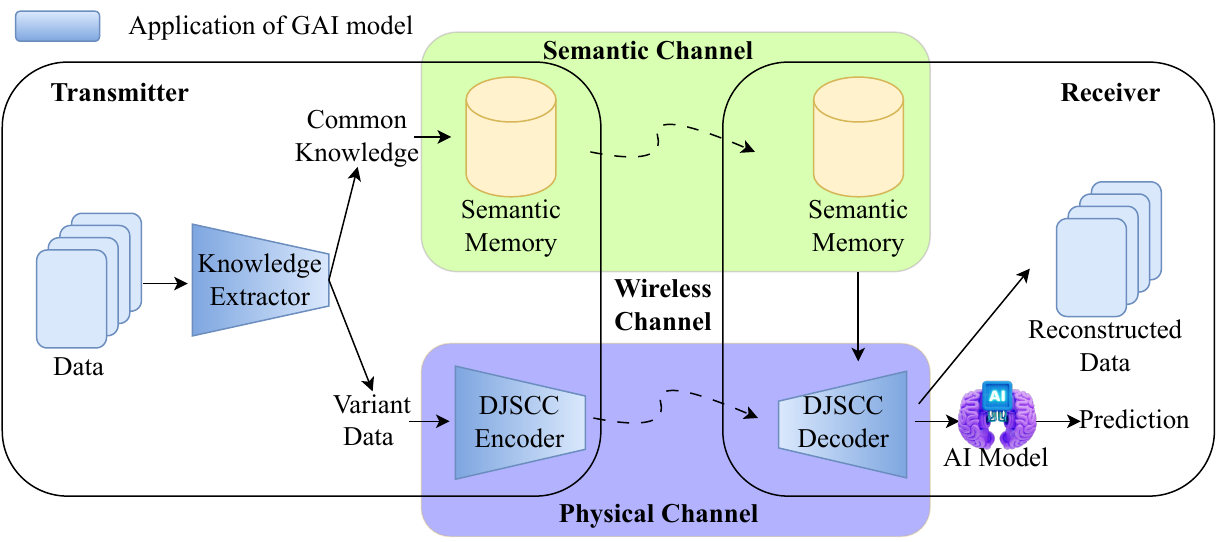}
   \caption{Illustration of GAI in semantic communication. The GAI take place in three blocks, i.e., Knowledge Extractor, DJSCC Encoder and Decoder. The Knowledge Extractor aims to generate both common knowledge and variant data from the original data. The DJSCC aim to compress and reconstruct the data with a high compression ratio and robust to noise.}
   \label{fig:GAI-KSC-SemCom}
\end{figure}

{\color{black} SemCom revolutionizes communication by considering the semantics of transmitted data to allow imperfect decoding while fulfilling the intended purpose and reducing unnecessary data traffic and energy consumption. However, SemCom also imposes three challenges: 1) generating common knowledge, 2) exploiting shared knowledge for efficient encoding and decoding, and 3) facilitating communication-free knowledge sharing among clients in distributed learning.
Following that, we first discuss how to generate common knowledge in SemCom. Next, we describe the strategies for integrating this common knowledge into data compression and reconstruction design. Afterward, we explore the role of GAI in accelerating the execution of AI-driven tasks and recommender systems in distributed networking, such as a knowledge generator and domain-invariant representation learning across devices. Fig.~\ref{fig:GAI-KSC-SemCom} illustrates a basic block of GAI in SemCom.
Finally, we conclude this section with useful lessons and key takeaways.}

\subsection{Knowledge Abstraction}\label{sec:knowledge-abstraction}
Knowledge abstraction is a technique used to represent abstract knowledge among users and acquire shared knowledge from distributed data across device domains. However, manually constructing KGs is labor-intensive and impractical for scalability \cite{Miller1995Nov, Speer2017Feb}. Toward automatic KG construction, three FL-enabled GAI designs in wireless networks have been recently proposed \cite{huang2023federated}, with three key knowledge techniques.

\subsubsection{Knowledge Summarization} 

Summarization techniques create concise abstracts that retain essential information for logical reasoning, omitting less crucial details. This results in the prevalence of leveraging Language Models (LMs) to extract knowledge from KGs. However, automatically learning optimal prompts typically requires abundant training data \cite{Lester2021Apr, Zhong2021Apr, Qin2021Apr}, which may not have useful meaning to new relationships. Despite its success with GPT-3, the approach in~\cite{West2021Oct} is not transferable to other LMs due to its reliance on ad hoc bounded learning and the scalability of GPT-3. 

To address the limitations of the aforementioned methods, BertNet \cite{Hao2022Jun} proposes an unsupervised learning approach to automatically generate diverse alternate prompts while requiring the input of minimal relationship definitions. Compared to conventional approaches that rely on extensive human-annotated data or large KGs, BertNet enables extraction of knowledge about new relationships previously inaccessible. 

\subsubsection{Knowledge Representation Optimization} 

This approach enhances knowledge representation efficiency by transforming complex rule-based structures into more compact formats, such as ontologies or Binary Decision Diagrams (BDDs). A previous study ~\cite{Pavlovic2022Jun} introduces a spatial-functional embedding for KG completion that represents entity pairs as points and relations as hyper-parallelograms in $\mathbb{R}^{2d}$. This enables pattern representation through the spatial relationship of hyper-parallelograms, providing a clear geometric interpretation of embeddings and captured patterns. 

However, creating representations for relation extraction in a KG remains challenging due to significant label noise in distantly labeled datasets. To tackle this circumvent, a solution called LERP was proposed~\cite{Ding2021Mar}. LERP is an efficient framework capable of learning prototypes for each relation based on contextual information, capturing the intrinsic semantics of these relations by observing noisy labels. On the other hand, summarizing KGs also encounter two critical problems: limited observations for model training and evolving characteristics of novel entities in temporal KGs. This motivates the presence of MetaTKGR in \cite{Wang2022Dec}, which employs meta-learning to adjust domain adaptation over time.

\subsubsection{Knowledge Aggregation} 
\label{sec:knowledge-aggregation}


This technique refers to the combination of similar and related knowledge from diverse data sets or learned models to reduce redundancy and enhance decision-making or the adaptive ability to new domains \cite{Wang2020Nov}. Nevertheless, exploiting automatic knowledge abstraction via entity retrieval encounters three primary obstacles: 1) Context and entity affinity primarily rely on vector dot products, possibly overlooking fine-grained interactions; 2) Large entity sets require a significant memory footprint to store dense representations; and 3) Training necessitates subsampling a suitably challenging set of negative data.

Driven by these challenges, a solution called Generative ENtity REtrieval (GENRE) is introduced \cite{DeCao2020Oct}, which combines a transformer-based generative model and a pre-trained NLP model. GENRE offers implementations with constrained decoding, ensuring that every generated entity belongs to a predefined candidate set. As a consequence, the outputs of the GAI model are less likely to deviate from the desired entities.

\subsection{Data Compression}\label{sec:data-compression}
\begin{figure}[t]
  \centering \includegraphics[width=\linewidth]{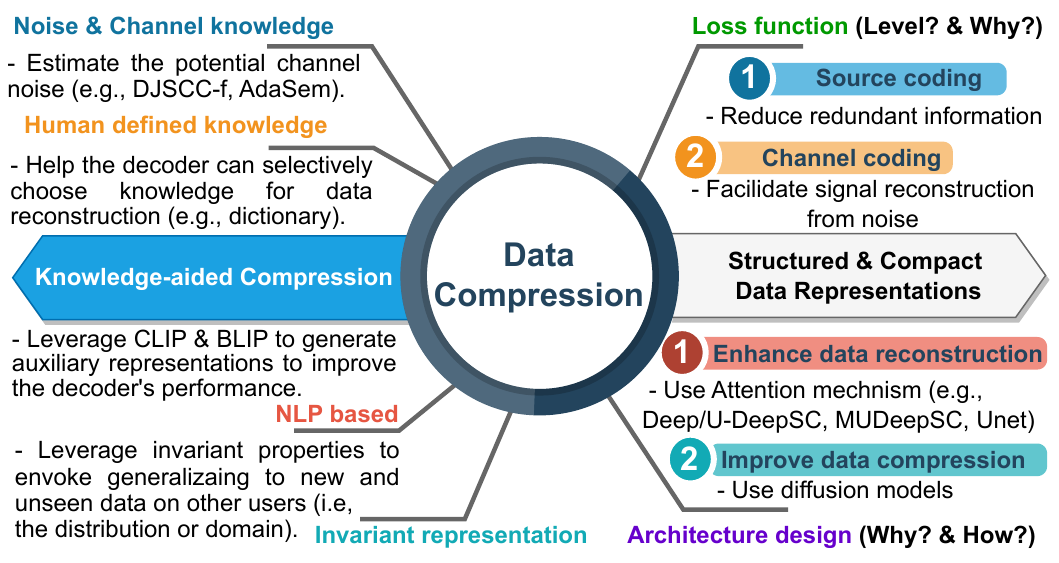}
   \caption{\color{black}Taxonomy of data compression techniques through GAI.}
   \label{fig:datacom}
\end{figure}

Data compression is vital in SemCom, particularly for communication with limited resources. The use of lossy compression is believed to be beneficial in improving data compression performance but at the expense of data transmission. With the emergence of SemCom and Goal-Oriented Semantic Communication (GOSC), the shortcomings of lossy compression have been overcome. To achieve this, two main objectives were addressed: improving data representation structure and implementing knowledge-aided techniques. Fig.~\ref{fig:datacom} illustrates the classification of data compression techniques using GAI, potentially strengthening mobile networking robustness.

\subsubsection{Structured and Compact Data Representation}
In DL and data representation, efficient data compression relies on compact and structured data representation~\cite{Ma2022Sep, wright2022high}. Recent studies have explored methods in two main categories: optimizing loss functions and designing model architecture.

\textbf{Loss Function Design.} This effort focuses on formulating an efficient loss function to minimize discrepancies between the original and reconstructed data while capturing noise features from the wireless channel. However, revisiting the classical Shanon theorem \cite{Shannon1948Jul} indicates that, for a memoryless communication channel, optimal communication involves two distinct processes as the data length approaches infinity:
\begin{enumerate}
    \item \textit{Source Coding}: Maximize compression to eliminate redundant information from images.
    \item \textit{Channel Coding}: Employ error-correcting codes to not only facilitate signal reconstruction in the presence of noise but also protect information through redundancy.
\end{enumerate}
While the separation theorem has been demonstrated to be effective in coding algorithms, it encounters two limitations. First, there is no such thing as an infinite number of bits. The overall distortion of signals depends not only on source information but also channel coding errors with finite block 
length, hindering bit allocation optimization and code design challenges. Second, achieving maximum-likelihood decoding, generally NP-hard, requires impractical computational power, hindering real-world implementation. 

Therefore, recent efforts focus on leveraging DL and GAI to JSCC system with acceptable approximation outcomes. Neural Error Correcting and Source Trimming (NECST) is the first effort in this cutting-edge area~\cite{Choi2019May}. This advanced DL framework compresses and corrects input images within a set bit-length constraint. NECST surpasses traditional models by integrating a discrete channel simulation, injecting noise directly into latent representations to enhance resilience in challenging channel conditions. When applied to the Deep JSCC (DJSCC) system~\cite{xu2023deep}, NECST activates self-learning, swiftly adapting to new environments without significant computational overheads from training anew.
The work in \cite{Tung2022Dec} tackles the challenge of JSCC over a discrete-input Additive White Gaussian Noise (AWGN) channel by mapping input images to codewords and decoders to minimize average end-to-end distortion. The specific problem of the end-to-end JSCC is regarded by a quantizer in the middle, where features are extracted from input images and quantized to constellation points through a DNN-based encoder. After transmission, the receiver reconstructs the image from noisy observations using another DNN. The study~\cite{Lee2023Apr} focuses on mitigating data noise from an impaired channel by capturing noise statistics and applying noise reduction based on channel behavior. The work in \cite{Zhang2021Dec} introduces a GAN-based approach for improved data reconstruction in lossy compression. Through a regularization function, this method can address both universal rate distribution and perception functions concurrently.

Error-Bounded Lossy Compression (EBLC) \cite{Liu2022Jul} is one critical technique for big scientific data processing that tailors compression according to the precision requirement of scientific datasets and ensures data fidelity within acceptable error bounds. However, to adapt to the highly accurate scientific analysis in diverse conditions, existing EBLC frameworks are mostly limited by configuration issues. To overcome this drawback, the work in  \cite{Underwood2022Mar} introduces a novel framework named OptZConfig to improve accuracy and speed, while the work in \cite{Mahmood2023Feb} develops universal rate distortion-perception representations to enhance data reconstruction.

With the demands of short training time and lower computational power, achieving high compression ratios without compromising data reconstruction, especially in 3D-based AI tasks like point cloud representations, is not a straightforward task. Inspired by this, DeepPCAC, a novel approach that directly encodes and decodes point cloud attributes, is introduced \cite{Sheng2021Jun} to surpass traditional compressed methods due to adopting geometry instead of using voxelization or point projection. 

As mentioned earlier, the advent of GOSC signifies a new era in lossy compression within SemCom due to no need for reconstructing the entire dataset. Instead, it compresses incoming data as long as AI task performance is achieved. This approach is effective as the AI era enters a prosperous phase, with machine-to-machine and human-to-machine communication becoming more prevalent in IoT Networks~\cite{Yang2023Nov}. The main feature of GOSC is to extract invariant features among domains~\cite{WangExplor2022} or classes~\cite{Wu2022Jan}. Therefore, as long as the encoders can capture invariant features, they identify crucial AI task features. Based on foundational principles, the authors in \cite{Dubois2021Dec}  integrate invariant learning in a lossy compressor, preserving lossless prediction in AI classification. This method encapsulates essential features for AI tasks, enabling lossy compression while maintaining consistent AI performance.

\textbf{Architecture Design.} There are two directions for this regard. The first focus is to design an efficient model architecture to enhance data reconstruction efficiency. 
Based on the Attention mechanism \cite{vaswani2017attention}, Deep SemCom (DeepSC) architecture for speech recognition applications is able to address noise and distortion challenges through joint semantic-channel coding \cite{Xie2021Apr}. Similarly, the upgrade versions, U-DeepSC and MUDeepSC \cite{Zhang2024Feb, Xie2022Jul}, accommodate text and image transmission and facilitate semantic information extraction. In that, exploiting a layer-wise knowledge transfer technique enhances information exchange across tasks, similar to UNet architectures \cite{Ronneberger2015Nov, Huang2020UNet, Isensee2021Feb}, ensuring information preservation.

The second focus is to design an efficient model architecture for data compression and reconstruction. The use of diffusion models \cite{Chang2023Jun} helps deduce the latent structure of a dataset by capturing how data points propagate through their latent space~\cite{Song2020Nov}. Upon conceptualizing the SemCom process as a sequential Gaussian model~\cite{Nguyen2023Sep}, the work in \cite{Theis2022Jun} introduces an improved lossy compression method capable of seamlessly integrating diffusion models. This method utilizes noise from the SemCom process's communication channel to improve compression and semantic encoding,  enhancing communication efficiency and data reconstruction noticeably.

\subsubsection{Knowledge-assisted Compression} 

Knowledge-Assisted Compression (KAC) is a recent groundbreaking compression approach that inherits from knowledge-based reasoning, as mentioned in Section~\ref{sec:knowledge-aggregation}. Knowledge-assisted compression approaches tackle the process of compressing or compacting knowledge representations to improve reasoning efficiency and storage requirements. KAC systems use explicit knowledge, often represented as rules, facts, or ontologies, to conclude, make inferences, or solve problems. Thus, the compressor and reconstructor can further improve the compression efficiency with the common knowledge among data.

\textbf{Noise and Channel Knowledge.} To address the oversight of semantic noise and system robustness in data encoding and compression, two efficient methods of Masked Autoencoder (MAE) \cite{Qiyu2022} and Vector Quantized VAE (VQ-VAE) \cite{Hu2023Apr} are introduced, where GAI is exploited to generate noise and pre-learning adversarial risks before transmission. Thus, both achieve robustness against semantic noise and enhanced semantic reconstruction via knowledge-based encoder-decoder architectures simultaneously. The work in~\cite{Zhang2023Aug} presents a quantization-based knowledge extractor, where an automatic mask generation based on semantic knowledge information is developed to regulate the compression ratio of transmitted data. By doing so, the approach enables the SemCom system to dynamically adapt to variable-length coding systems. To further improve compression efficiency within the SemCom system introduced in~\cite{Xie2021Apr}, the authors of~\cite{Xie2023Jun} employ LSTM techniques to formulate a memory-aided SemCom system. With historical information about the data and compression efficiency, this approach yields higher compression efficiency without compromising communication efficiency, meaning there is no need to transmit additional data via the communication channel. {\color{black}Another approach is to leverage the noise and channel feedback from the receivers to perform estimation of the ideal data transmission at the encoder to mitigate the channel noise \cite{2022-Compress-DJSCCQ,2024-Compress-Commin,2024-Compress-DeepJSCC-MIMO,2024-Compress-GenerativeJSCC,2024-SemCom-AdaSem,2023-SemCom-GILA}.}

{\color{black}\textbf{Human-defined Knowledge.} The authors in \cite{2023-Compress-SCDA} introduced one of the earliest approaches to incorporating language-based knowledge into data compression. This study proposes a framework wherein a predefined library of language knowledge, such as an electronic dictionary, is accessible. The decompression process can then selectively utilize relevant knowledge from this library to enhance compression efficiency significantly. Thus, this approach not only underscores the potential of leveraging language-based knowledge for data reconstruction but also achieves a relatively simple design.}

{\color{black}\textbf{NLP-based Knowledge Generator.} This approach can be divided into two folds. The first approach utilizes pretrained LLMs to generate auxiliary representations that enhance the performance of the decoder \cite{2024-SemCom-LLM-Graph, 2023-SemCom-SIAC}, with two advantages. LLMs trained on extensive datasets produce representations that encapsulate a wealth of knowledge and no additional training is required, making the system easy to integrate.
The second approach is to leverage prompting from the model of BLIP and CLIP to improve the decoder performance \cite{2024-SemCom-LaMoSC, 2024-SemCom-LoC}. 
This facilitates huge computation performance for inference time without overparameterization typical in LLMs.}

{\color{black}\textbf{Causality-Invariant Representation.}
This approach enables practical causal reasoning in SemCom, recently attracting considerable attention \cite{2024-SemCom-SurveyData}. The generalized characteristics of causality invariance benefit radio nodes in learning and establishing a stable representation for a specific semantic content element \cite{2022-DG-CIRL, 2021-DG-CausalMatching, 2024-FL-FedCDH}.  Once this representation is formed, radio nodes can use it consistently to describe semantic content without being affected by variations in distribution, domain, or context. Given causality invariant representations, causality-based neural symbolic AI can indeed help achieve minimal representations, create symmetric communication channels \cite{2024-SemCOm-NeSy}, enable generalizability, and reduce data transmitted. For example, SemCom frameworks in \cite{2024-SemCOm-NeSy, Thomas2022} can achieve reliable communication with minimal data transfer across diverse services and uses. With an emphasis on GFlowNets, such GAI-inspired networks serve as the basis means in highlighting task distributions with changing contexts to identify task event transitions and form a graph for GNN to improve data reconstruction.
Similarly, using a causal SemCom method enhances communication efficiency among distributed clients significantly \cite{Thomas2023Nov}.} In that, GAI models with causally invariant representation abilities help the SemCom system prioritize the transmission of the most meaningful features to the goal-oriented task at the recipient's end. Another example is employing a knowledge-enhanced SemCom framework~\cite{Wang2023May} to help the receiver actively utilize the data in the knowledge base for semantic reasoning and decoding. This is achieved using a transformer-based knowledge extractor to find relevant factual triples for the received noisy signal. 

\subsection{Generative AI in Distributed Learning}\label{sec:distributed-learning}
\begin{figure}[t!]
  \centering \includegraphics[width=\linewidth]{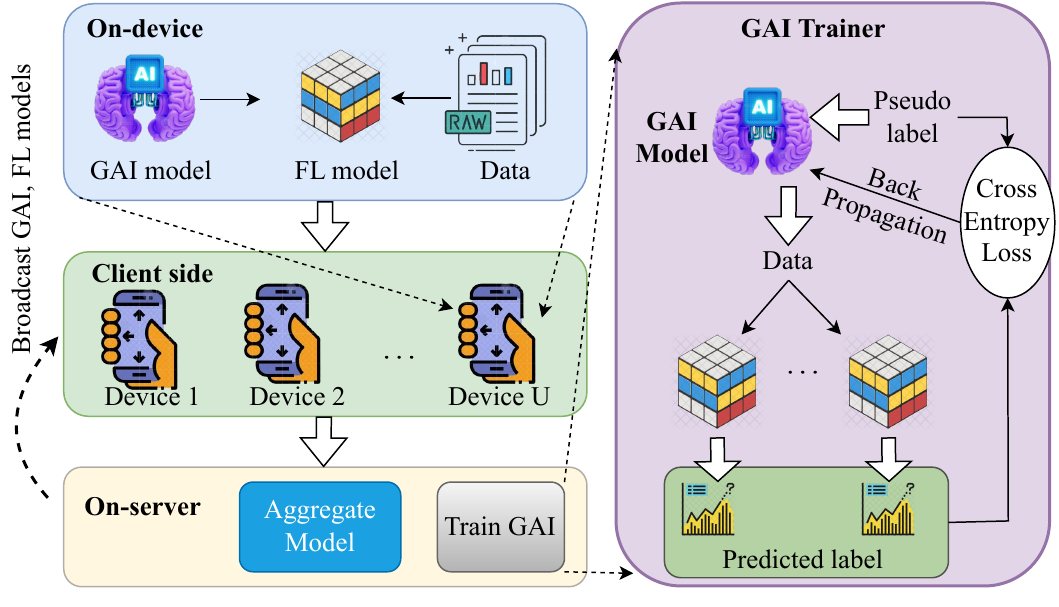}
   \caption{Architecture of integration of GAI into FedGen.}
   \label{fig:FedGen}
\end{figure}

GAI has demonstrated remarkable robustness in enhancing performance in distributed learning. To be specific, GAI techniques produce generative data representations that extract generalized knowledge or meaningful features, supporting training on distributed local devices without requiring access to unseen data domains. In FedGen~\cite{Zhu2021Jul}, the server based on local model performance cooperatively trains a GAI model as a lightweight generator to amass user information and that is capable of generating pseudo data representation without accessing their true data (see Fig.~\ref{fig:FedGen}). Thus, each local user can have the additional knowledge of other distributed clients, without sharing data among distributed devices. This method is robust as it can significantly enhance the performance and communication efficiency of FL, all while ensuring that user data remain secure and private. In \cite{du2023generative}, an AaaS framework based on AIGC models is designed to address the challenges in training and deployment overheads for edge networks.

\begin{figure}[t]
  \centering \includegraphics[width=\linewidth]{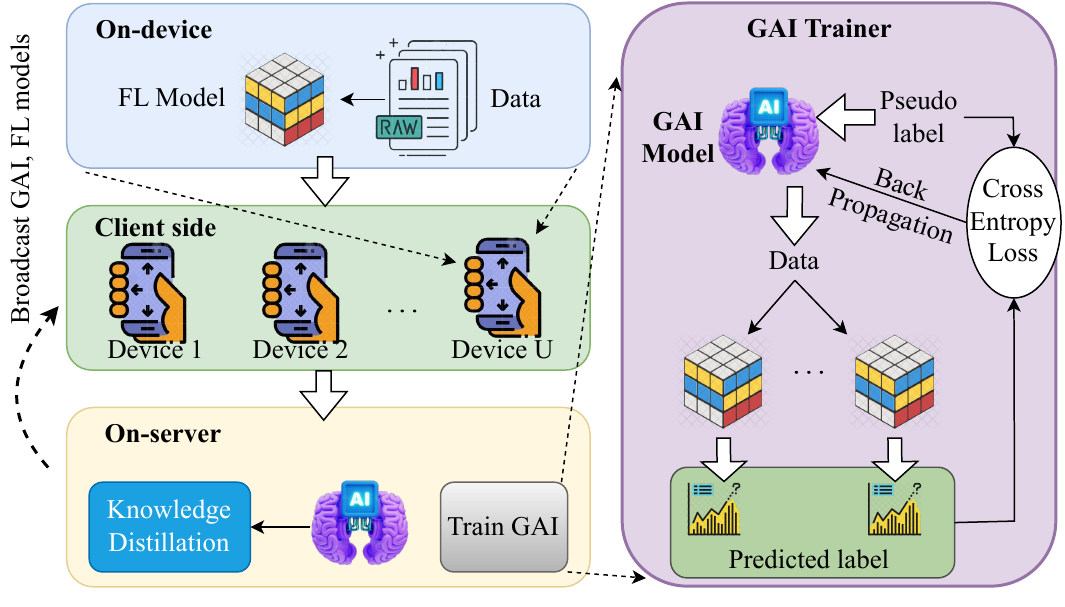}
   \caption{Architecture of integration of GAI into DENSE.}
   \label{fig:Dense}
\end{figure}

Similar to FedGen, which leverages GAI to train a GAI model, DENSE~\cite{Zhang2022Dec} designs a GAN-based data generation network that enables the generation of on-server data. In contrast to FedGen, where GAI is employed for on-server data generation, DENSE employs a multi-teacher knowledge distillation technique (see Fig.~\ref{fig:Dense}). This technique ensure that the generated data accurately captures the entire system's data distribution and characteristics. The local models collaboratively instruct a student, denoted as the global model, inheriting knowledge from all teachers. Furthermore, this approach can be enhanced by incorporating Label-Driven Knowledge Distillation (LKD)~\cite{Nguyen2022Sep}. This methodology enables the student to selectively learn the most meaningful knowledge from each teacher, significantly improving distillation efficiency compared to other conventional knowledge distillation approaches. Consequently, it is feasible to apply the ensemble learning on the server, boosting the FL efficiency while retaining the communication efficiency.

Unlike other GAN approaches, the work in~\cite{Nguyen2022Jul} adopts a distinct approach by treating model parameters as data. Specifically, the authors introduce HCFL, a technique capable of compressing and reconstructing model parameters. It enables the FL system to circumvent substantial communication overheads. In particular, applying HCFL to FL systems can maintain stable accuracy when the distortion rate of the aggregated model decreases proportionally with network size.

\begin{figure}[t]
  \centering \includegraphics[width=0.85\linewidth]{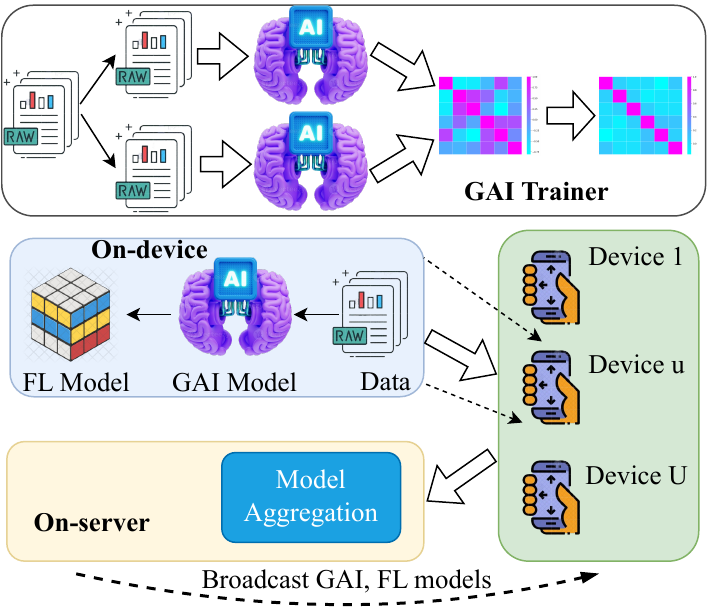}
   \caption{Architecture of integration of GAI into FCCL.}
   \label{fig:FCCL}
\end{figure}

Aside from the aforementioned methods, invariant learning is a unique approach to enhancing learning on local devices in distributed learning. In invariant learning, users collect their local invariant representations or cross-domain invariant representations by sharing the most consistent format of data (invariant features) across devices. This way, it minimizes the need for extensive communication resources, a representative form of knowledge-aided SemCom in distributed learning. For instance, Federated Cross-Correlation and Continual Learning (FCCL) \cite{Huang2022} and FCCL+ \cite{Huang2023Oct} are innovative methods inspired by Barlow Twins principles, which help identify invariants across domains in the FL system (See Fig.~\ref{fig:FCCL}). As an alternative method to discover invariant features, the authors in~\cite{Huang2023Wenke} focus on class-wise invariance by introducing cluster prototypes of contrastive learning. Unlike \cite{Huang2022,Huang2023Oct,Huang2023Wenke}, FedPAC employs a structured causal model to acquire goal-oriented invariant representations, demonstrating robustness in learning personalized representations for each client without extensive knowledge sharing among devices~\cite{Xu2023Jun}.

\begin{table*}[!th]
\centering
\caption{GAI for Optimizing Semantic Communication Scenarios.}
\label{tab:GAI_SemCom}
\resizebox{\textwidth}{!}{
\begin{tabular}
{|p{0.5cm}|p{0.65cm}|p{1.15cm}|p{2.5cm}|p{3.0cm}|p{1.1cm}|p{3.5cm}|p{4.2cm}|}
\hline
\textbf{Apps}               
& \textbf{Ref} 
& \textbf{Model}
& \textbf{Inputs}
& \textbf{Outputs} 
& \textbf{Loss function}
& \textbf{Performance} 
& \textbf{\textcolor{black}{{Limitations}}} \\ \hline
\multirow{9}{*}{\rotatebox[origin=r]{90}{Knowledge Abstraction}}  &   \multirow{2}*{\cite{DeCao2020Oct}}        &    Encoder-Decoder    &   Textual input source   &    Summarized knowledge    &    Softmax   & Competitive results on document retrieval with less memory  & \textcolor{black}{Required pre-processing of training data, and only applicable to text-based data applications} \\ \cline{2-8} 
 &      \cite{Han2023May}        &   Logical Network             &   Knowledge Graph with missing edges &   Contextual information as vectors   &  Gradient-based                      &  Outperforms other rule learning and estimates expressive logical rules                    & \textcolor{black}{Hard to deploy and fine-tune LERP for various networking uses} \\ \cline{2-8} 
 &   \cite{2024-SemCom-AdaSem} & \textcolor{black}{VAE} & \textcolor{black}{Video and motion sensor measurements} & \textcolor{black}{Localization Data}  & \textcolor{black}{VIB + APP} & \textcolor{black}{Effectiveness in localization task-oriented data} & \textcolor{black}{Caused straggling due to channel feedback requirements} \\ \cline{2-8} 
 &  ~\cite{zou2023wireless}            &    LLMs            &    Data and information from edge networks             &  Collaborative planning and decision-making by multi-agent LLMs                 &     Q-reward loss       &   Demonstrates the potential benefits of on-device LLMs to solve network intents.  & \textcolor{black}{The impractical algorithm due to the overwhelming computation overheads} \\ \cline{2-8} 
 &  ~\cite{Erdemir2023Jun}            &   StyleGAN             &  Distorted reconstructions of images               &  Improved perceptual quality of transmitted images                 &  MSE +  GAN loss               & Improves perceptual quality by performing denoising using style-based GAN                    & \textcolor{black}{High training complexity, mode collapse, and slow convergence rate} \\ \hline
\multirow{22.5}{*}{\rotatebox[origin=r]{90}{Data Compression}} &   ~\cite{Hu2023Apr}           &    \textcolor{black}{Vision Transformer}            &   \textcolor{black}{Image input source}   &   \textcolor{black}{Improved quantized vector with noise mask} & \textcolor{black}{Cross-entropy}                     & \textcolor{black}{Generate simulated semantic noise to combat channel noise via adversarial training}    & \textcolor{black}{Not suitable for IoT frameworks due to the modeling over-parameterizations} \\ \cline{2-8} 
    &  ~\cite{Yazdinejad2023Feb}            &    AE-LSTM            &  Heterogeneous data from IoT devices               &  Identification of out-of-norm activities for cyber threat hunting in IIoT                &        MSE                &    Ensemble model outperforms conventional ML with better compression                 & \textcolor{black}{Not applicable to different types of datasets} \\ \cline{2-8} 
    &  ~\cite{du2023yolo}            &   Transformer and Diffusion                  &  Captured images from edge devices               &  Semantic optimized resource allocation extracted from images                &      MSE                  &   Reduces transmission costs by extracting important semantic information & \textcolor{black}{Not suitable for IoT systems due to the overparameterized network} \\ \cline{2-8} 
    &  ~\cite{Lee2023Apr}            &    Diffusion model            & Heterogeneous data from IoT devices                &  Improved quality reconstructed images                 &   Log-likelihood                   &    Significant gain in data reconstruction under different settings of channel distortion                   & \textcolor{black}{Required large test times and computation overheads due to the intensive diffusion process} \\ \cline{2-8}
    &  ~\cite{Xie2021Apr}            &    Transformer            & Captured images from edge devices                &  Realistic image reconstructed based on semantic segmentation                &   BLEU                   &    Immense resource gain in bandwidth reduction during semantic segment transmission  & \textcolor{black}{Not feasible for IoT systems due to transformer architecture, high computation, and large model size } \\ \cline{2-8}
    &  ~\cite{Mentzer2020}            &    RC        & Captured images from edge devices          &  Lossless reconstructed images.                &   Cross-entropy               &    Lossless image reconstructed from lossy compressed data & \textcolor{black}{Unable to simulate information loss of LBP at different scales} \\ \cline{2-8}
    &  ~\cite{Thomas2023Nov}            &    Causal semantic communication            & Sequential signal                &  Reconstructed sequential signal                &   KL Divergence                   &    Improves signal reconstruction using causal theorem to predict the potential signal to support the reconstruction                   & \textcolor{black}{Information loss due to neglected non-causal components} \\ \cline{2-8}
     &  ~\cite{2024-SemCOm-NeSy, Thomas2022}            &    NeSy            & Graph representation                &  Knowledge based on graph                &   KL Divergence                   &    Neural-symbolism to improve the AI task of reconstructing knowledge from the graph & \textcolor{black}{Straggling due to the requirement to receive feedback from the receiver to improve the data compression} \\ \hline
\multirow{5}{*}{\rotatebox[origin=r]{90}{Semantic Resource Allocation}}  &  ~\cite{Zhang2024Apr}            &   DRL             &  Real-time data from vehicular networks               &   Improved service quality, reliable guidance to receiving vehicles               &     MSE                   &     Enhances service quality, reliability, and efficiency in GAI-enabled vehicular networks                 & \textcolor{black}{Required large test times and computation overheads due to the intensive diffusion process} \\ \cline{2-8} 
    &  ~\cite{lin2023unified}            &  ISGC              &   User inputs for semantics and graphics rendering              &   Optimized resource allocation and high-quality content generation               &   Variational autoencoder                     &   Optimal resource allocation strategy for the Metaverse                 & \textcolor{black}{Required large test times and computation overheads due to the intensive diffusion process} \\ \cline{2-8} 
    & ~\cite{du2023exploring}             &   Diffusion model     &  Wireless network collaboration tasks              &    Optimized execution of AIGC tasks              &  MSE                      & Optimized edge computational resource utilization                      & \textcolor{black}{Required large test times and computation overheads due to the intensive diffusion process} \\ \cline{2-8} 
    &  ~\cite{park2022enabling}            &    RSA            &    Multi-modal data from human/machine agents             &  Locally rendered scenes and interactions in the Metaverse                &  MSE                      &   Utilizes distributed learning and multi-agent RL for efficient communication & \textcolor{black}{High communication overheads due to frequently shared semantic representations among agents} \\ \hline
\multirow{5}{*}{\rotatebox[origin=r]{90}{Distributed Learning}}     
    &  ~\cite{zou2023wireless}            &   LLMs             &  Edge networks and multi-agent system parameters               & Collaboratively planned and solved tasks in wireless networks                   &    Q-reward loss      &   Self-governed networks with on-device LLMs for intelligent decision-making                   & \textcolor{black}{The impractical algorithm due to overwhelming computation overheads} \\ \cline{2-8} 
    & ~\cite{huang2023federated}             &  Diffusion models + LLM     &  AIGC tasks, settings of wireless networks, and privacy               &  Diverse, personalized, and high-quality content generation                &    Adversarial loss  & Improved learning efficiency, privacy protection, and content quality      & \textcolor{black}{Not suitable for IoT systems due to fine-tuning foundation models} \\ \cline{2-8} 
    &   \cite{Zhu2021Jul}    &  GAN              &  Heterogeneous user data               &  Ensembled knowledge for guiding local models                &       KL Divergence               &  Improved generalization performance in FL with minimum communication resource & \textcolor{black}{Poor generalization in learning due to estimating the models' representations from noise (at the server)} \\ \cline{2-8} 
    &  ~\cite{huang2023ai}            &  Diffusion model         &     Network scale, environment parameters, and user demands            &  Customized network solutions and expert-free optimization                &     Custom ensemble loss                   & Best adaptation to objectives and scenarios and creation of novel designs &
    \textcolor{black}{Required large test times and computation overheads due to the intensive diffusion process} \\ \hline
\end{tabular}
}
\end{table*}

\subsection{Lessons Learned and Key Takeaways}
GAI plays a crucial role in the development of SemCom, especially in generating data suitable for various subject-specific requirements. Consequently,  applying GAI to various subdomains brings several benefits to SemComs, including creating more compact and structured data representations, significantly improving data compression efficiency in SemCom, and enhancing communication efficiency.

Furthermore, the robust development of GAI, particularly in summarization and knowledge extraction, enables the system to extract common knowledge from data. In other words, identifying elements that are domain-invariant or label-invariant (consistent regardless of changes in data) allows the system to leverage such invariant information multiple times without passing it through the physical communication channel. This leads to a substantial reduction in communication overhead without diminishing the amount of transmitted information.

{\color{black}The key takeaways from this section are as follows.
\begin{itemize}
    \item \textbf{Key 1.} GAI markedly enhances compression performance via the creation of concise and organized data representations.

    \item \textbf{Key 2.} GAI empowers the SemCom capability to abstract knowledge effectively, facilitating high-performance data compression without sacrificing information.

    \item \textbf{Key 3.} GAI facilitates the generation of pseudo data on distributed devices, allowing these devices to adapt to the domains of other devices in distributed learning scenarios.

    \item \textbf{Key 4.} GAI can assist the encoder in generating invariant features from data that contribute significantly to AI performance. This improvement leads to enhanced learning while concurrently achieving a substantial reduction in communication costs.
\end{itemize}
}

\section{Challenges and Future Directions}

\label{Sec:challenges}
GAI has demonstrated its effective abilities in network management, wireless security, and SemCom for mobile and wireless networking developments. However, integrating GAI into modern networks also raises new issues that must be addressed. {\color{black}In what follows, we have explored some viable solutions for the issues of complexity of GAI models in Section~\ref{complexity}; scalability in Section~\ref{Inter-Scala}; security, privacy, and trust in Section~\ref{Secu-Priva-Trus}; and standardization in Section~\ref{standardizations}.
Finally, we summarize all critical issues, the respective solutions, and the raised research directions in Table~\ref{tab:challenges}.}

\begin{table*}[!th]
	\centering
	\caption{Summary of Challenges, Respective Solutions, and Future Research Directions of GAI in Mobile and Wireless Networking.}
	\label{tab:challenges}
    {\color{black}\begin{tabular}{|p{1.7cm}|p{3cm}|p{6cm}|p{5.8cm}|}
        \hline
        \textbf{Challenges} & \textbf{Description} & \textbf{Possible Solution} & \textbf{Future Research Directions} \\
        \hline
        \multirow{6}{*}{Complexity} 
        &
        \textbullet~High training and testing time\par 
        \textbullet~Overparameterization of inputs 
        & 
        \textbullet~\textit{Lightweight features}: Lightweight tuning paradigm \cite{ChenXiang2021Aug}, distributed GAI networking solutions like innovative generative IoV architecture \cite{XieApr2024} or online generative FL model \cite{Zhang2022Jul}. \par  
        \textbullet~\textit{Effective learning approaches}: Finite-difference score matching \cite{NEURIPS2020_de6b1cf3}. \par 
        
        & 
        Future research in GAI-integrated networks should prioritize developing GAI in forms of distributed networking like generative FL or generative collaboration in FL. Additionally, the design of onboard-integrated tiny/lightweight GAI models is also important for compatibility with IoT devices.  
        \\ \hline
        \multirow{7}{*}{Scalability} 
        &
        \textbullet~Transferring model\par 
        \textbullet~Division of infrastructure networking
        & 
        \textbullet~\textit{Customizable attributes}: Unsupervised learning BertNet approach coupled with lightweight tuning features \cite{Hao2022Jun,ChenXiang2021Aug}. \par 
        
        \textbullet~\textit{Flexible network control paradigm}: hybrid federated and centralized learning \cite{Elbir2022Jun},  knowledge-defined network model \cite{Ashraf2022May}, virtual network function \cite{Passas2022Oct}, and multi-modal GAI paradigm \cite{BrodimasJuly2024}. \par  
                
        & Future research in GAI-integrated networks should focus on developing GAI modules with customizable options to be compatible with diverse network components. At the same time, there is a need for a designed network control paradigm with GAI plugging functions. \\ \hline
        \multirow{6}{*}{\begin{tabular}[l]{@{}l@{}}Security,\\ Privacy, \\and Trust\end{tabular}} 
        &
        \textbullet~Heterogeneity of network components \par 
        \textbullet~Differential security protocols\par
        \textbullet~Consumer's perspective on GAI use\par
        & 
        \textbullet~\textit{Enhancing Data Privacy and Integrity}: Adoption of physical countermeasures \cite{Aouedi2024Jul} or defence strategies against the threats of GAI \cite{Mavikumbure} and developing GAI-based supply chain security and blockchain security mechanisms  \cite{Sai2024Apr}. \par 

        \textbullet~\textit{Robust Single/Multilayered Protocols}: Using an automatic generation of distributed GAI algorithms \cite{VazJun2023} and defence-in-depth by leveraging layers of controls protected by a hardware-enhanced trusted execution environment \cite{MofidiFeb2024}. \par 

        \textbullet~\textit{Building Trustworthy in GAI Interactions}: Evaluating AI trustworthiness \cite{Shen2022Feb} or transforming `\textit{black box}' GAI models into more transparent and understandable `\textit{glass box}' models \cite{Franzoni2023Jun}. \par 
                
        & Future research in GAI-based networks should continuously improve security and privacy mechanisms at the physical-layer transmission while developing GAI frameworks oriented towards supply chain and blockchain securities. Besides, it is also required to design automatic distributed GAI algorithms combined with multilayered security protocols and realize GAI performance frameworks for evaluating trustworthiness in GAI-based systems. \\ \hline

        \multirow{6}{*}{Standardization} 
        &
        \textbullet~GAI model evaluation \par 
        \textbullet~Rule and policy in use, operation, and maintenance \par
        & 
        \textbullet~\textit{GAI model design}: Using a comprehensive GAI evaluation framework \cite{zhao2024lightweight}. \par 
        
        \textbullet~\textit{Standardization bodies}: Relying on 3GPP standardization \cite{TR-28.908,TS-28.105,TR-37.817} for the 5G core network (OAM and RAN) and enhanced 5G NR air interface \cite{TR-38.843}. 
        & Future research should prioritize creating a general GAI framework, defining the mechanisms needed, and exploring new use cases for GAI. These efforts lay the foundation for 6G, which will integrate GAI and communication as key features. \\ \hline

    \end{tabular}}
\end{table*}

\subsection{Complexity}\label{complexity}

{\color{black}Integrating GAI into mobile and wireless networks can significantly benefit various advanced applications and services. However, GAI-based LMs typically require large computational resources and abundant data for training and fine-tuning \cite{Lester2021Apr, Zhong2021Apr, Qin2021Apr}. For instance, ControlNet \cite{2023-DM-ControlNet} requires a significant amount of training time from scratch until the model is well performed. 
Specifically, the model is trained for 300 GPU hours using NVIDIA A100 80GB GPUs, taking user scribbles as input conditions, and for 400 GPU hours using the same GPUs for the semantic segmentation task. Additionally, the pre-trained GAI-based LMs are not always available for specific tasks, causing a significant growth in the training time with the number of task requests. Moreover, compared to traditional DNNs, the inference duration (i.e., testing time) in GAI-based diffusion models is considerably longer due to the progressive denoising process. Evidently, the DDPM \cite{2020-DM-DDPM} requires up to 1,000 iterative steps to denoise a single data sample, with each step taking approximately 10 seconds. Importantly, GAI-based models are often constructed with highly complex architectures. For instance, ChatGPT-3, a text generation tool, requires 175 billion parameters, while StableDiffusion v1.5 \cite{Rombach_2022_CVPR} uses 1 billion ones (6.04 GB) for both UNet and CLIP, which serve as image and text encoders. 

To overcome the aforementioned challenges of IoT infrastructure networks, where IoT components do not always have sufficient capabilities (i.e., size, memory, and computational requirements) to effectively perform GAI-based LM processes (e.g., unmanned aerial vehicles \cite{2024-SemCom-LLM-Graph}), GAI-integrated framework should be lightweight. In this regard, a lightweight tuning paradigm in \cite{ChenXiang2021Aug} can be adopted to minimize the fundamental tasks of labeling data for training GAI models while achieving flexible re-modulation of attention and adaptation of pre-trained weights, making GAI modules more flexible for low-resource scenarios and enhancing their ability to transfer knowledge across IoT domains. In addition, exploiting distributed networking solutions for GAI is also crucial for reducing computation and storage tasks. For example, one can generalize the innovative generative Internet-of-Vehicles (IoV) architecture to establish a collaborative fine-tuning mechanism for pretrained GAI models, along with a self-adaptive harmony search-based resource allocation strategy~\cite{XieApr2024}. 
This framework allows resource-limited IoT devices to achieve rapid, customized, and lightweight GAI, by optimizing time and energy consumption through collaborative fine-tuning. Another solution is to consider online generative FL, as in \cite{Zhang2022Jul}, to deal with non-independent and identically distributed (non-IID) trajectory data across data owners in travel time estimation applications. Specifically, the framework establishes the generative global model shared by all clients to infer real-time road traffic conditions and adjusts a personalized model for each client to analyze their driving habits, thereby correcting the residual errors from the global model's localized predictions. 
Moreover, designing effective learning approaches is also needed, such as the finite-difference score matching approach in \cite{NEURIPS2020_de6b1cf3}, for ML applications to parallelly approximate any-order directional derivative, without gradient computation during training. 
In summary, computational requirements and high complexity hugely affect the feasibility of GAI-based IoT solutions. Thus, it is expected that further efforts will be made in these suggested directions.}

\subsection{Scalability}\label{Inter-Scala} 

{\color{black}GAI is typically trained on specific LMs (e.g., CLIP \cite{2021-GAI-CLIP}, BLIP \cite{2022-GAI-BLIP}) or specific tasks (e.g., ControlNet \cite{2023-DM-ControlNet}) with constrained learning properties. 
The scalability of GAI in transferring one domain to others is quite limited in the context of IoT \cite{West2021Oct}, where IoT data often exhibit varying characteristics.
Besides, the division of networking infrastructures into centralized and decentralized components, such as computing architecture, data models, and application protocols, further complicates the seamless integration of GAI into existing mobile and wireless networks \cite{QiaoSep2020}. 
Moreover, to flexibly adapt to varying requirements from service providers, GAI-based frameworks must incorporate high-level features that enable them to specify the desired network capabilities.

From the mentioned above, future works should first focus on developing GAI models with strong learning capabilities, which are less affected by classification tasks, such as the unsupervised learning approach of BertNet in \cite{Hao2022Jun}, coupled with a lightweight tuning paradigm in \cite{ChenXiang2021Aug}. 
To accommodate both centralized and decentralized networking deployments, developing hybrid GAI models that leverage the advantages of distributed and centralized learning \cite{Elbir2022Jun} is a promising approach. This means that only clients with sufficient computational power and resources will perform FL, while the clients with limited capability will directly transfer their data to the centralized server for processing centralized training tasks. 
Besides, to provide intelligent network services with real-time decision-making capabilities, GAI-based frameworks for SDN must have robust data processing and quick response features. For example, applying the knowledge-defined network modelling approach in \cite{Ashraf2022May} can build a self-driving SDN environment with better precision and efficient performance under variable traffic loads and delays, even with disparate routing schemes in a flat topology. Meanwhile, using the virtual network function solution in \cite{Passas2022Oct}, which enables dynamic adaptation and reconfiguration of orchestrator deployments based on demand, can proactively achieve autoscaling functions in a cloud-native network. To achieve high dynamicity in interpreting the user's intent and ensure compatibility with various orchestrators or new operating systems, exploiting the multi-modal GAI paradigm in \cite{BrodimasJuly2024} is also a promising approach.}

\subsection{Security, Privacy, and Trust}\label{Secu-Priva-Trus}

{\color{black}The inherent diversity and complexity of emerging IoT applications and services (e.g., intelligent transportation, smart homes/cities, healthcare, financial sectors, academia, Industry 5.0, and military aspects like joint radar and communication) have led to several challenges to GAI integration, where any IoT component, from data to devices, can be a potential vulnerability point, including GAI-deployed components. 
Notably, when a GAI-based security module is poisoned, it can become a powerful tool for attackers, combined with other attacking approaches \cite{Aouedi2024Jul}, to manipulate/extract sensitive information without resistance \cite{du2023spear}. 
In addition, each IoT element or group is operated and controlled by different security protocols (e.g., sensor-to-sensor, sensor-to-IoT node, mobile-to-sensor, etc.), presenting unique challenges. These factors introduce diverse security risks and create considerable gaps between security measures. 
Below are possible strategies for these risks.

\textbf{Enhancing Data Privacy and Integrity.} To enhance user privacy, future works should first focus on constructing countermeasures from the physical-layer transmission perspectives, such as clustering-based methods, differential privacy, encryption and homomorphic encryption, generative steganography strategies, and secure multi-party computation \cite{Aouedi2024Jul}. Moreover, the adoption of defence strategies against the threats of GAI is also needed to enhance the personal or confidential data from the potential privacy loss and risks during data exchange and aggregation, such as red and blue teaming to identify system vulnerabilities, adversarial training to strengthen GAI models, explainable AI to interpret trained model outputs, security awareness and training to address the risks associated with the misuse of GAI, and input/output filtering to identify, mitigate, and potentially reject suspicious results \cite{Mavikumbure}. To ensure the data integrity, it is required to implement security mechanisms such as authentication, authorization, and continuous monitoring. For example, developing GAI-based supply chain security approaches can create digital maps of supply chain routes, locations, and intermediaries for tracking the journey of goods, identify potential vulnerabilities, and exploit blockchain security mechanisms for creating digital signatures or hashes and for monitoring blockchain network activities and transactions to detect unusual patterns or behaviours \cite{Sai2024Apr}.

\textbf{Robust Single/Multilayered Protocols.} Unlike classical security, the evolving nature of cyber threats continuously adapts to AI variations. Consequently, GAI-secured systems must automate and proactively update their measurement mechanisms to protect against emerging vulnerabilities \cite{Andreoni2024Aug}. For example, an automatic generation of distributed GAI algorithms in \cite{VazJun2023} can be adopted for automating the process of fault-tolerant distributed algorithms, such as reliable broadcast, total order broadcast, causal broadcast, and consensus. Moreover, developing multi-layered or defense-in-depth strategies is crucial to ensure that if one protective layer is breached, the other layers continue to offer protection. An example of a multi-layered strategy includes a firewall as the first line of defense, an IDS as the second line, and regular vulnerability assessments as the third line. Another example is to apply the principle of defence in depth by leveraging layers of controls protected by a hardware-enhanced trusted execution environment \cite{MofidiFeb2024}.

\textbf{Building Trustworthy in GAI Interactions.} The traditional design of ML and AI models has focused solely on achieving accurate output predictions. However, the recent shift towards generative IoT is evident in the emergence of generative search and recommendation systems capable of content retrieval, reuse, and creation to meet user needs. Given GAI's significant impacts on science, health, and humanity (i.e.,  ethical and legal concerns), the development of future intelligent systems must focus on achieving accurate results while ensuring transparency, fairness, and unbiasedness. This consequently raises the following important questions:
\begin{itemize}
    \item  What does it mean to trust a GAI, and how can we evaluate its trustworthiness?
    \item  What are the mechanisms/tools for building trustworthy GAI, and what is the role of interpretable GAI in trust?
\end{itemize}
To what extent these questions significantly enhance trust in GAI-based systems has not yet been answered. However, a similar AI approach is referenced in \cite{Shen2022Feb}. This involves first distinguishing human-GAI trust from human-GAI-human trust in conjunction with evaluating a GAI’s contractual trustworthiness. Next, using a ladder of model access helps define the role of interpretability in trust. Finally, creating behaviour certificates that are more accurate, relevant, and understandable, rather than merely opening up black boxes, is essential. Another feasible solution is to transform `\textit{black box}' GAI models into more transparent and understandable `\textit{glass box}' models \cite{Franzoni2023Jun}. This approach highlights the clear importance of transparency for stakeholders (i.e., users, developers, and policymakers) as well as the trade-offs between transparency and other objectives, such as accuracy and privacy.}

\subsection{Standardizations}\label{standardizations}

{\color{black}GAI's mobile and wireless networking integration is a promising strategy for the development of advanced 5G and 6G networks. Moreover, factors like the complexity in LLM for training, data fine-tuning, incompatibility in infrastructure networks, communication protocols, and control models, as well as ethical and legal issues, hinder GAI applications. Since GAI is in its early stages of evolution for various purposes, standardizing its involved learning model/architecture/commercial modules remains an unanswered question. For example, how can we define lightweight GAI models? Answering this question requires a comprehensive GAI evaluation framework \cite{zhao2024lightweight}, with rigorous sequential steps from dataset constructions, feature extractors (types of learning models), and score calculators, to the methods for model evaluation and enhancement. Another question is how to integrate GAI into mobile and wireless networks and operate it. This is an extreme challenge to standardization bodies (e.g., IEEE, ITU, and 3GPP) and big techs (e.g., KT Corp, Samsung, and Cisco). As one of the first aspirants in this direction, 3GPP has recently engaged in preliminary AI/ML initiatives for the 5G core network, with emphasis on two sectors: the operations, administration, and maintenance (OAM) and the radio access network (RAN), through 3GPP documents TR 28.908, TS 28.105, and TR 37.817 \cite{TR-28.908,TS-28.105,TR-37.817}. To clarify the potential of AI/ML-based algorithms in enhancing the 5G New Radio (NR) air interface, 3GPP has recently released the latest version of 3GPP TR 38.843 \cite{TR-38.843}.}

\section{Conclusion}
\label{sec:Conclusion}
In this article, we comprehensively reviewed the fundamentals and potential applications of GAI for the future Internet and mobile networking. We first provided the evolution and principles of GAI and then discussed its applications for important research issues in mobile communications and networking, including network management, wireless security, and SemCom. In each application, we provided lessons learned and key takeaways from the reviewed literature to help researchers develop research into applied GAI for mobile networking. As research on GAI for mobile communications and networking is in the early stages, we need to consider both opportunities and challenges toward the realization of GAI in this research area. As such, we finally discussed key challenges and potential research directions, such as technical challenges, privacy, security, trust, ethical and legal concerns, and open-source frameworks and tools.



\begin{IEEEbiography}
[{\includegraphics[width=1in,height=1.25in,clip,keepaspectratio]{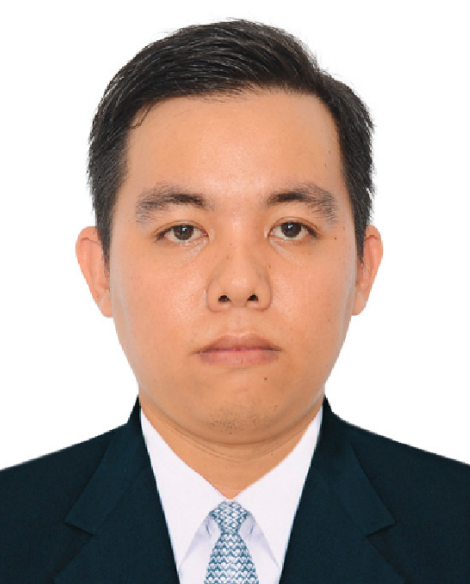}}]
 {Thai-Hoc Vu} (M'24) received the B.S. degree in Electrical and Telecommunication Engineering from the Posts and Telecommunications Institute of Technology, Ho Chi Minh City, Vietnam, in 2018. From 2018 to 2020, he was a Research and Development Engineer with the Research and Development, Strategy Department, at Saigontourist Cable Television Company Limited, Ho Chi Minh City. Currently, he is a Ph.D. candidate in the Department of Electrical, Electronic and Computer Engineering, University of Ulsan (UOU), Republic of Korea. He has authored and co-authored more than 30 transactions/journals and 7 flagship conference papers. He was also a co-recipient and recipient of the IEEE International Conference on Communications and Electronics (ICCE) Best Paper Student Awards in 2022 and 2024, respectively, and the Vietnamese Young Scientists in Korea Award from the Vietnamese Students Association in Korea and the Vietnam Embassy in the Republic of Korea. Also, he received the prestigious Exemplary Reviewer Certificate of IEEE Communications Letters in 2023. He acts as a reviewer for major international IEEE venues. His research interests lie in the areas of wireless communications, communication theory, new radio access, and machine learning for future wireless communication.
 \end{IEEEbiography}
 \begin{IEEEbiography}
 [{\includegraphics[width=1in,height=1.25in,clip,keepaspectratio]{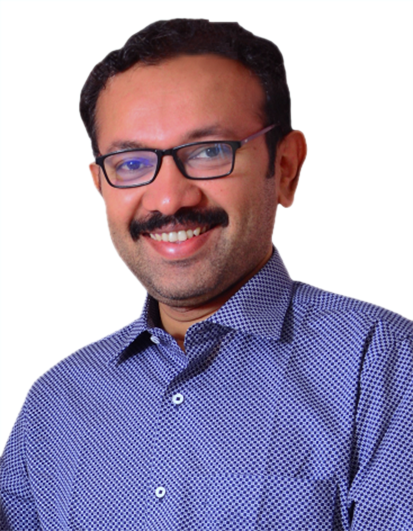}}]
 {Senthil Kumar Jagatheesaperumal}
 received his B.E. degree in Electronics and Communication Engineering from Madurai Kamaraj University, Madurai, India in 2003. He received his post-graduation degree in Communication Systems from Anna University, Chennai, India in 2005. He received his Ph.D. in Information and Communication Engineering from Anna University, Chennai, India in 2017. He is an Associate Professor (Senior Grade) in the Department of Electronics and Communication Engineering, Mepco Schlenk Engineering College, Sivakasi, India. To his credit, he was granted four utility patents to date. He received two funded research projects from National Instruments, USA, each worth USD 50,000, during the years 2015 and 2016. He also received another funded research project from IITM-RUTAG in 2017 worth Rs.3.97 Lakhs. Recently, in 2024, he received a sponsored research project from IHub-Data worth Rs.8.2 Lakhs. His area of research includes the Internet of Things, Robotics, Embedded Systems, and Wireless Communication. He is a Member of IEEE, IETE and ISTE.
 \end{IEEEbiography}
\begin{IEEEbiography}
[{\includegraphics[width=1in,height=1.25in,clip,keepaspectratio]{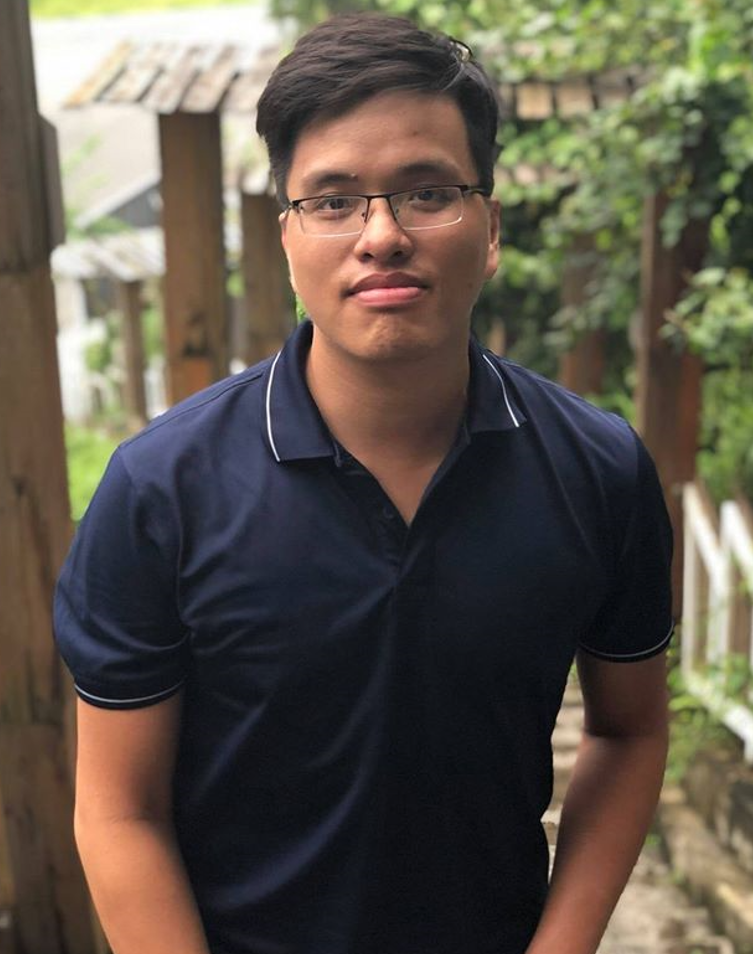}}]
 {Minh-Duong~Nguyen} received a B.S. degree in electronics and telecommunications engineering from the Hanoi University of Science and Technology, Vietnam, in 2016. He was a DSP engineer in Viettel R\&D Center from 2017 to 2019. He was working in electronic warfare (i.e., electronic warfare systems and GPS/4G jamming systems). He was a senior embedded engineer in Vinsmart, Vingroup, from 2019 to 2020, developing physical layers for the 5G Base Station. He is currently pursuing his Ph.D. degree in the Department of Information Convergence Engineering at Pusan National University, South Korea. He is currently serving as a reviewer for journals and conferences, e.g., IEEE TNNLS, IEEE TWC, IEEE JSTSP, IEEE JSAC, IEEE IoTJ, ICML, and ICLR. He was nominated as the top reviewer of ICML 2022.
 His research interests include reinforcement learning, federated learning, representation learning, multi-task learning, meta learning, and SemCom.
 \end{IEEEbiography}
 \begin{IEEEbiography}
 [{\includegraphics[width=1in,height=1.25in,clip,keepaspectratio]{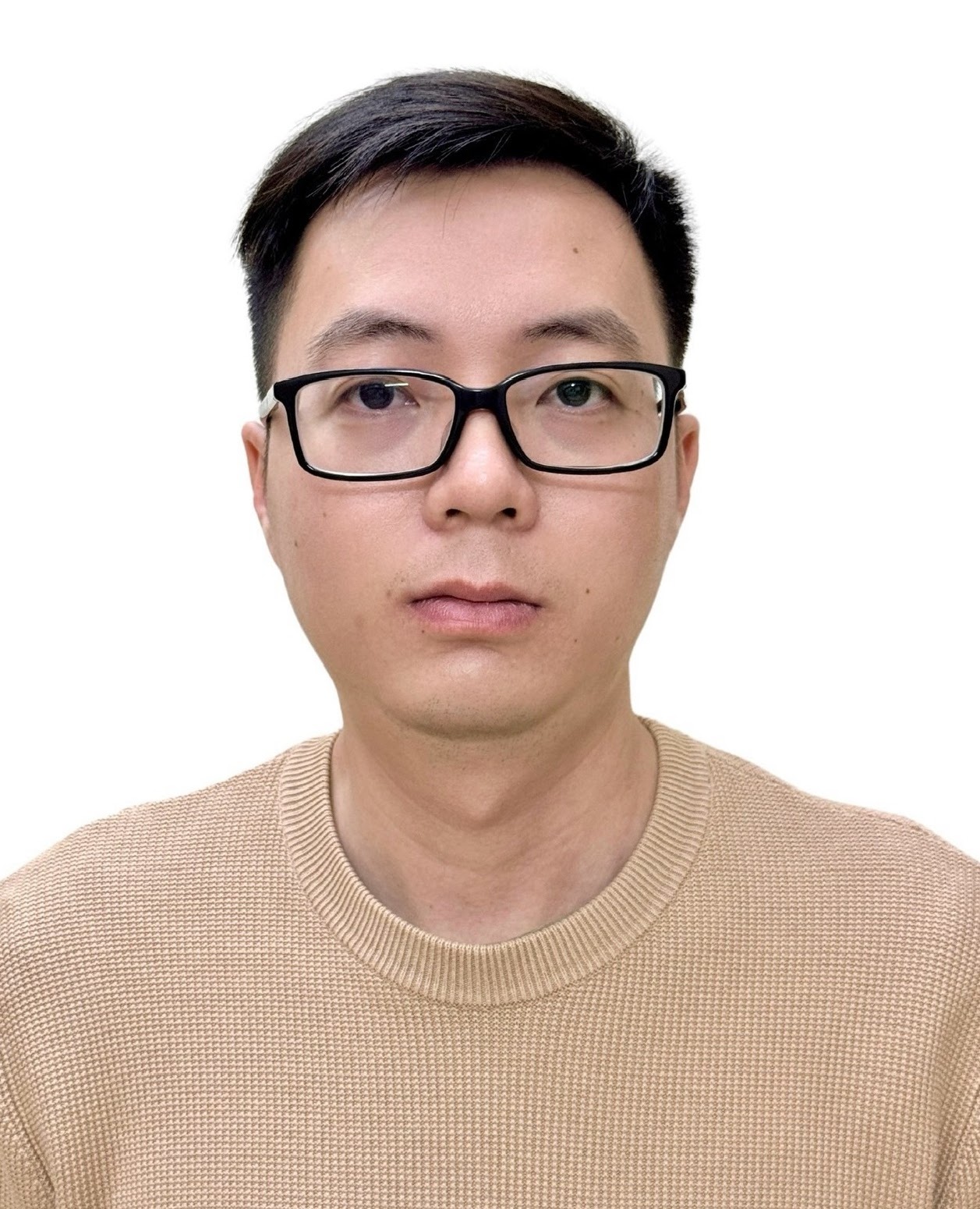}}]
 {Nguyen Van Huynh} Dr. Huynh Nguyen received the PhD degree in Electrical and Computer Engineering from the University of Technology Sydney (UTS), Australia in 2021. He is currently a Lecturer at the Department of Electrical Engineering and Electronics, University of Liverpool (UoL), United Kingdom. Before joining UoL, he was a Postdoctoral Research Associate in the Department of Electrical and Electronic Engineering, Imperial College London, United Kingdom. His research interests include cybersecurity, 5G/6G, IoT, and machine learning. He received several awards from Google, UTS, and IEEE ComSoc, including Google PhD Fellowship, UTS FEIT Research Excellence Award, IEEE WCL Exemplary Reviewer, and IEEE GLOBECOM and IEEE ICC Travel Grants. He has co-organized several IEEE conferences in communications and networking (e.g., IEEE GLOBECOM, IEEE ICC, and IEEE FNWF) as TPC members and co-chairs. He is an associate editor for IEEE Transactions on Vehicular Technology and was a guest editor for IEEE Transactions on Cognitive Communications and Networking. Dr. Huynh Nguyen has been listed among the most cited researchers by Scopus and Stanford University since 2022.
 \end{IEEEbiography}

 \begin{IEEEbiography}
 [{\includegraphics[width=1in,height=1.25in,clip,keepaspectratio]{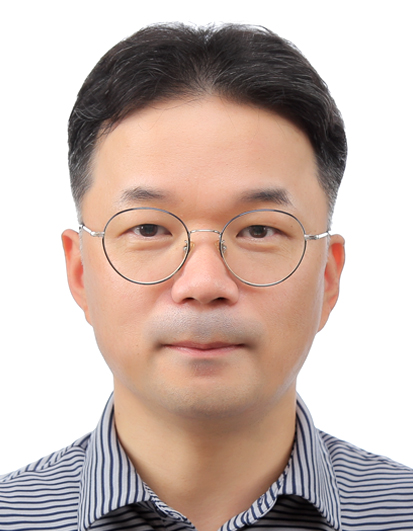}}]
 {Sunghwan Kim}  (Senior Member, IEEE) received the B.S., M.S., and Ph.D. degrees from Seoul National University, Seoul, South Korea, in 1999, 2001, and 2005, respectively. From 2005 to 2007, he was a Postdoctoral Visitor with Georgia Institute of Technology, Atlanta, GA, USA. From 2007 to 2011, he was a Senior Engineer with Samsung Electronics, Suwon, South Korea. From 2011 to 2024, he was a Professor with the Department of Electrical, Electronic and Computer Engineering, University of Ulsan, Ulsan, South Korea. He is currently a Professor with the School of Electronic Engineering, Kyonggi University, Suwon, South Korea. His research interests include 5G/6G communications, IoT communications, multiple access, deep-learning, transformer, error correction codes, and DNA-based storage.
 \end{IEEEbiography}
\begin{IEEEbiography}
[{\includegraphics[width=1in,height=1.25in,clip,keepaspectratio]{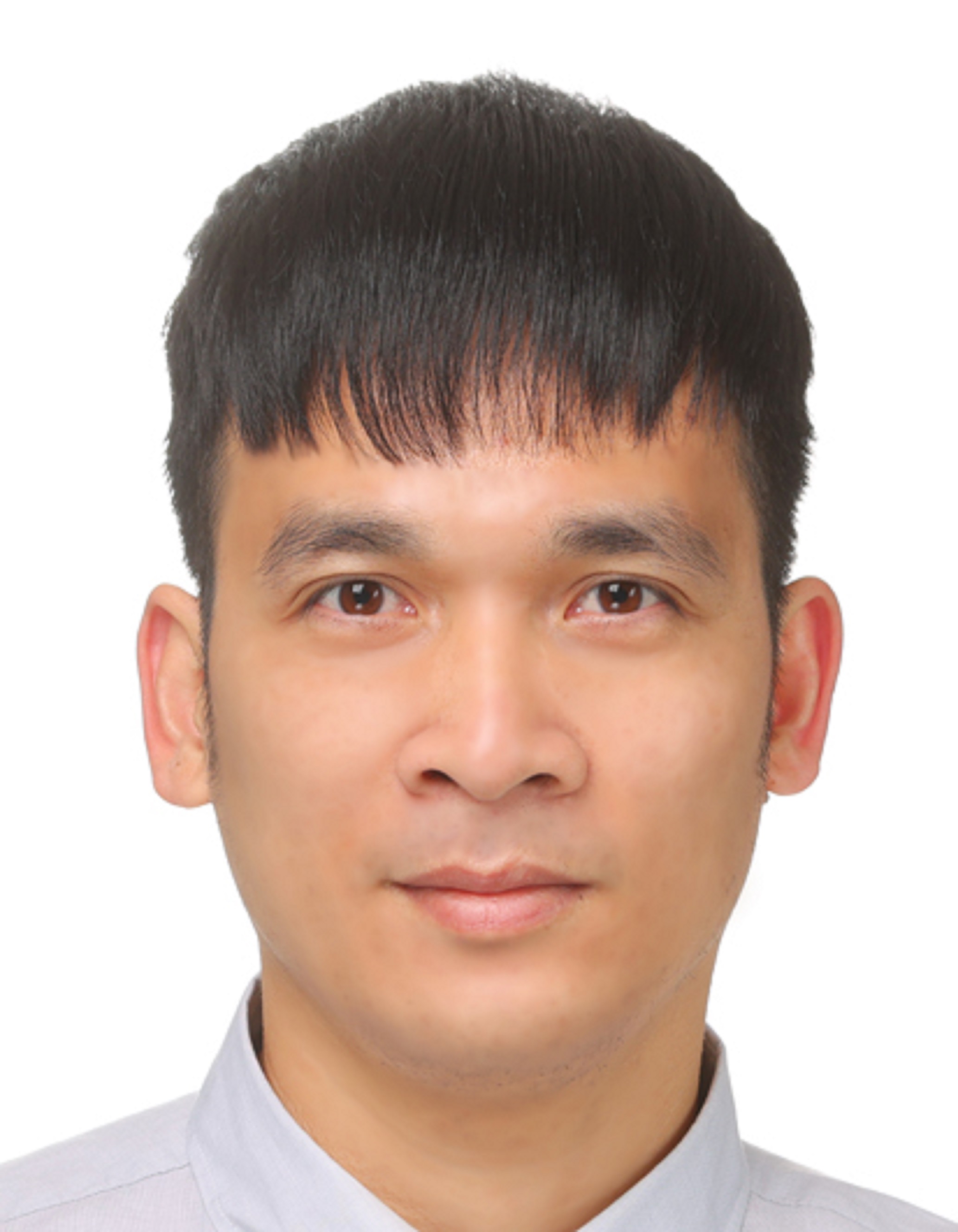}}]
{Quoc-Viet Pham} (M'18, SM'23) is currently an Assistant Professor in Networks and Distributed Systems at the School of Computer Science and Statistics, Trinity College Dublin, Ireland. He earned his Ph.D. degree in Telecommunications Engineering from Inje University, Korea, in 2017.
 
He specialises in applying convex optimisation, game theory, and machine learning to analyse and optimise cloud edge computing, wireless networks, and IoT systems. He was awarded the Korea NRF funding for outstanding young researchers for the term 2019-2024. He was a recipient of the Best Ph.D. Dissertation Award in 2017, Top Reviewer Award from IEEE Transactions on Vehicular Technology in 2020, Golden Globe Award in Science and Technology for Younger Researchers in Vietnam in 2021, IEEE ATC Best Paper Award in 2022, and IEEE MCE Best Paper Award in 2023. He was honoured with the IEEE ComSoc Best Young Researcher Award for EMEA 2023 in recognition of his research activities for the benefit of the Society. He is currently serving as an Editor for IEEE Communications Letters, IEEE Communications Standards Magazine, IEEE Communications Surveys \& Tutorials, Journal of Network and Computer Applications, and REV Journal on Electronics and Communications. He served as a (Guest) Editor for IEEE Internet of Things Journal, IEEE Internet of Things Magazine, IEEE Transactions on Consumer Electronics, Computer Communications, and Scientific Reports.

 \end{IEEEbiography}

\end{document}